\documentclass[12pt,preprint]{aastex}
\usepackage{amssymb, amsmath}
\usepackage{color}
\usepackage{morefloats}
\usepackage{hyperref}
\usepackage{natbib}

\shortauthors{Neugent et al.}

\begin{document}

\title{A Modern Search for Wolf-Rayet Stars in the Magellanic Clouds. IV. A Final Census \altaffilmark{*}}

\author{Kathryn F.\ Neugent\altaffilmark{1,2}, Philip Massey\altaffilmark{1,3}, and Nidia Morrell\altaffilmark{4}}

\altaffiltext{*}{This paper includes data gathered with the 1-m Swope and 6.5-m Magellan Telescopes located at Las Campanas Observatory, Chile.}
\altaffiltext{1}{Lowell Observatory, 1400 W Mars Hill Road, Flagstaff, AZ 86001; kneugent@lowell.edu; phil.massey@lowell.edu}
\altaffiltext{2}{Department of Astronomy, University of Washington, Seattle, WA, 98195}
\altaffiltext{3}{Department of Physics and Astronomy, Northern Arizona University, Flagstaff, AZ, 86011-6010}
\altaffiltext{4}{Las Campanas Observatory, Carnegie Observatories, Casilla 601, La Serena, Chile; nmorrell@lco.cl}

\begin{abstract}
We summarize the results of our four year survey searching for Wolf-Rayet (WR) stars in the Large Magellanic Cloud (LMC) and Small Magellanic Cloud (SMC). Over the course of this survey we've discovered 15 new WRs and 12 Of-type stars. In this last year we discovered two rare Of-type stars: an O6.5f?p and an O6nfp in addition to the two new Of?p stars discovered in our first year and the three Onfp stars discovered in our second and third years. However, even more exciting was our discovery of a new type of WR, ones we are calling WN3/O3s due to their spectroscopic signatures. We describe the completeness limits of our survey and demonstrate that we are sensitive to weak-lined WRs several magnitudes fainter than any we have discovered, arguing that there is not a population of fainter WRs waiting to be discovered.  We discuss the nature of the WN3/O3s, summarizing the results of our extensive spectroscopy and modeling.  We also examine the important claim made by others that the WN3/O3s are isolated compared to other massive stars.  We find that if we use a more complete sample of reference massive stars,  the WN3/O3s show the same spatial distribution as other early WNs, consistent with a common origin.  Finally,  we use this opportunity to present the ``Fifth Catalog of LMC Wolf-Rayet Stars," which includes revised coordinates and updated spectral types for all 154 known LMC WRs.
\end{abstract}


\section{Introduction}
Four years ago we set out to survey the entire optical disks of both the Large Magellanic Cloud (LMC) and Small Magellanic Cloud (SMC) in search of Wolf-Rayet (WR) stars. Our survey was spurred on by the recent discovery of a rare oxygen-rich WR \citep{NeugentWO} in one of the LMC's most visually striking OB associations, Luke-Hodge 41. At the time, the WR content of the LMC was thought to be relatively complete, so the discovery of such a star pointed to the possibility of a whole group of undiscovered LMC and SMC WRs. Thus we began a search for WRs that combined both image subtraction and photometry with spectroscopic followup with a hope of discovering new WRs.

Our survey has been remarkably successful. We have found a host of interesting emission lined objects including 15 new WRs, 12 Of-type stars and even a low-mass X-ray binary (LMXB). In this past observing season we found two very rare types of Of stars, an Onfp and on Of?p. Before this study there were only 3 known Of?ps in the Magellanic Clouds, and before this observing season our survey had already added 2 more, thus almost doubling the known contents of these rare O stars in the MCs. However, while these discoveries are exciting, we were most surprised by the spectra of 10 of our newly discovered WRs \citep{PaperI, PaperII, PaperIII}. These spectra are unlike any observed WR with both strong absorption and emission lines. We believe them to be single stars since they are too faint to be in O+WR binary systems and they can be modeled using a single set of physical parameters. Further discussion of these stars can be found in \citet{Neugent17}. Here we discuss our new results from the final season of our discovery observations, and summarize the results of our previous studies. We also provide an updated catalog of LMC WR stars in the Appendix.

In Section 2 we discuss our Observations and Reductions, in Section 3 we highlight our new discoveries, in Section 4 we comment on our completeness limits, in Section 5 we discuss the isolation of the newly found WN3/O3 stars, and finally we conclude in Section 6. We additionally present an updated catalog of LMC Wolf-Rayet stars in the appendix. No new WRs were discovered in the SMC and thus we do not include them in the appendix.

\section{Observations and Reductions}
We imaged the optical disk of the LMC and SMC using the Swope 1-m telescope on Las Campanas, Chile. Observation and reduction procedures can be found in \citet{PaperI, PaperII, PaperIII} (hereafter called Paper I, Paper II, and Paper III, respectively) but are briefly discussed here. Our exposure times were 300s through each of three 50 \AA\ wide interference filters. The first was centered on C\,{\sc iii} $\lambda4650$, a strong line found in WC stars. The second was centered on He\,{\sc ii} $\lambda 4686$, a strong line found in WN stars, and the third on the neighboring continuum ({\it CT}) centered at 4750 \AA\ (see Figure 3 in Paper I). A detailed description of the bias subtraction and flat fielding can be found in Papers II and III.

To discover WRs, we observed 537 fields over the course of 38 nights within the five and a half year period of 2013-09-21 to 2017-11-20 (UT). We surveyed all of our planned fields except for one that we missed in the outer regions of the SMC. Many of the fields were observed two or more times as sky conditions improved from night to night. The total area surveyed was 40.8 degrees$^2$ of the LMC and 27 degrees$^2$ of the SMC. These numbers are quite close to the 38.5 degrees$^2$ (LMC) and 28.5 degrees$^2$ (SMC) that were originally planned. Our field locations and overall coverage can be seen in Figure~\ref{fig:coverage}.

For each field, we subtracted the {\it CT} image from the {\it WN} and {\it WC} images using the {\it High Order Transform of PSF And Template Subtraction (HOTPANTS)} code written by Andrew Becker and described briefly in \citet{Becker}. We then visually identified WR candidates (see Figure 4 in Paper I) using aperture photometry to identify those that were photometrically significant. Using the two techniques in tandem lead to an initial list of possible targets. We ran these targets through VizieR to exclude previously known emission-line sources (such as planetary nebulae) or red objects ($B-V>0.5$), which show up as candidates due to a strong absorption band falling within the continuum filter. 

With our WR candidate list in hand, we used the 6.5 m Magellan telescope and the Magellan Echellette (MagE) spectrograph for spectroscopic confirmation (we began the project on the Clay telescope before MagE moved over to the Baade for the following three years). Over the past four years we have had four nights of observing time for spectroscopic followup. For the last and final year of this project we were awarded two half nights, 2017 February 7 (UT) and 2017 February 8 (UT). The sky was clear and seeing was 0.7-0.9\arcsec\ for the three LMC stars and 1.7\arcsec\ for the one SMC star as reported on the guiding camera. Exposure times were between 600 to 1200s depending on the star's magnitude. Further reduction techniques are described in Papers I, II and III. 

\section{New Discoveries}
\subsection{Strange Of-type Stars}
As discussed above, WRs are discovered primarily by their strong emission lines, generally He\,{\sc ii} $\lambda 4686$. However, there are other classes of stars with He\,{\sc ii} $\lambda 4686$ in emission, just to a lesser extent. These stars are called ``Of" stars and during the last season of this survey we found two unusual ones, an O6.5f?p and an O6nfp. Their coordinates and photometry are shown in Table~\ref{tab:emission}.

In general, O stars are classified by the relative strengths of the He\,{\sc i} and He\,{\sc ii} absorption lines while the luminosity class is primarily determined by the presence or absence of the He\,{\sc ii} $\lambda 4686$ emission line. This line is sensitive to mass-loss which is in turn dependent upon luminosity and metallicity. Normal O stars are classified by comparing their spectra to those shown in \citet{WF} or \citet{Sota}. Of-type stars, such as the ones we discuss here, have both N\,{\sc iii} and He\,{\sc ii} emission as is expected for supergiants. However, the stars we have discovered are both rare types of Of stars. 

The first is an ``Of?p". We discovered and discussed two others of these in Paper I. The classification criteria come from \citet{WalbornOf?p} and describes Of-type stars whose C\,{\sc iii} $\lambda 4650$ emission is comparable in intensity to the N\,{\sc iii} $\lambda\lambda 4634,42$ emission. Of?p stars are also highly magnetic \citep{Munoz, Bagnulo2017} and are believed to be oblique magnetic rotators \citep{WalbornOf?p}. A spectrum of the discovered Of?p star is shown in Figure~\ref{fig:Of?p}. 

The second Of-type star we discovered is an Onfp. These stars are Of-type stars show a central reversal (absorption, in other words) in their He\,{\sc ii} $\lambda 4686$ emission line. We discovered one of these in Paper II and discuss the classification criteria there. We additionally presented two new Onfp stars in Paper III, LMCe078-2 and LMCe113-1. The type was introduced by \citet{Walborn73} and recently more were found by \citet{WalbornOnfp}. The latter claims these stars are typically in binary systems. The spectrum of the O6nfp is shown in Figure~\ref{fig:Onfp}. 

Both Of?p and Onfp-type stars are extremely rare. For example, there are only around 30 known Onfp stars in the Magellanic Clouds \citep{WalbornOnfp,Walborn14}. But, more surprisingly, before our survey there were 5 Of?p stars known in the Galaxy and only 3 in the Magellanic Clouds \citep{Walborn15}. We have doubled the known number of these stars in the Magellanic Clouds as part of this survey.

\subsection{Interesting Non-Emission Stars}
In constructing our candidate lists, false positives are to be expected. In some cases the reasons are easy to understand: very red stars have an absorption band in our continuum band, and thus show up as potential WRs. These are easily eliminated prior to spectroscopy by their broad-band colors, if available. In other cases, short-term variability can cause a star to be brighter in the emission-line filters than in the continuum filter. Some of the candidates we eliminated by spectroscopy had been picked up because of this. In other cases the reasons are less obvious.

We had two examples of this in our final year of the survey. One of these, SMCe090-1, was a known GALEX source, making it particularly attractive, but our first spectrum showed it lacked emission. We classify it as a B6~II star. Its radial velocity is consistent with membership in the Milky Way disk. Our second ``loser" is LMCe177-1, a high priority candidate whose spectrum revealed it is a faint A0 star, with an intermediate luminosity class. Its radial velocity is consistent with SMC membership, but that is unlikely given its magnitude. The radial velocity would not be peculiar for a halo giant, and this is what we expect it is.

\subsection{Revisiting LMC174-1, a More Highly Reddened WN3/O3}

In the original discovery paper of the WN3/O3s there was one star, LMC174-1, whose spectrum was sufficiently poorly exposed that we could only tell that it was a likely member of this class.  It was about a magnitude fainter than the others, and our somewhat noisy spectrum showed that it too had He\,{\sc ii} $\lambda$4686 and N\,{\sc v} emission with no sign of N\,{\sc iv}, and an absorption spectrum dominated by H and He\,{\sc ii}. But our S/N was too poor to be certain that He\,{\sc i} was not present.  Our first fluxed spectrum also indicated surprisingly high reddening, which we also wanted to check.

During a recent observing run on MagE on UT 2018 Feb 04 we obtained 3$\times$1200s exposures of this star under sub-arcsec conditions (0\farcs6) and perfectly clear skies.  The spectrum is shown in Fig.~\ref{fig:lmc1741}, compared to that of LMC079-1, a more typical example of our WN3/O3s.  The upper figure shows a small normalized section of the spectra with the principal lines identified, while the lower figure compares the spectral energy distributions.

It is apparent from the figure that LMC174-1 has slightly stronger emission, and that the absorption is somewhat weaker than that of LMC079-1.  To place this in better context, the spectrum of 8 of the WN3/O3s are almost identical.  Of the other two, LMCe055-1 shows a lower excitation emission and absorption spectrum, and is more accurately described as WN4/O4, and will be discussed in its own paper shortly.  However, of the nine that have a similar excitation spectrum (i.e., WN3, with no N\, {\sc iv}, but strong N\, {\sc v}), LMC174-1 is most different.  Although we would not call it intermediate between the WN3/O3s and the normal WN3s, its spectral appearance tends in that direction.

We also confirm here that it is also the most heavily reddened of any of the WN3/O3s, as is shown in the lower panel of Fig.~\ref{fig:lmc1741}, where we compare its spectral energy distribution to that of LMC079-1.   In Paper I we used the original fluxed discovery spectrum to estimate that LMC174-1 had an $A_V=1.6$ [$E(B-V)=0.5$], high compared to that of most other early-type stars in the LMC \citep{Massey07}, which have $A_V\sim0.4$ [$E(B-V)=0.13$]. As shown in Table~\ref{tab:WRs}, broad-band photometry of LMC174-1 gives $B-V=0.14$ \citep{Zaritsky04}. The other WN3/O3s have a median $B-V$ value of $-0.17$ (\citealt{Neugent17}, Table 2).  This implies that LMC174-1's $E(B-V)$ is $\sim$0.3 greater than most of the other WN3/O3s.  If we adopt an $E(B-V)$ value of $0.12$~mag as typical, then that $E(B-V)$ value of LMC174-1 is  0.43~mag, in good agreement with our original estimate.   Using our new spectrophotometry presented here, we measure a somewhat higher value, $E(B-V)=0.6$, or $A_V=1.9$. We adopt $A_V=1.6$ as a good compromise.  

It is tempting to suggest that these two facts are related, that the stronger emission/weaker absorption and the higher reddening are somehow connected, as they are both unique amongst the WN3/O3s, suggesting a significant local (circumstellar?) component to the reddening. Indeed, a close of examination of the spectrum of LMC174-1 shows that the Balmer absorption lines appear to be superposed upon broad emission that we do not see in LMC079-1. (See, e.g., H$\gamma$ in the upper panel of Fig.~\ref{fig:lmc1741}.)  Might this be indicative of a disk? If so, this could be contributing to the reddening.

The counter argument, however, is that the strength of the Ca\,{\sc ii} H and K interstellar lines, and in particular the prominent diffuse interstellar band (DIB) feature at $\lambda$4430, points to true interstellar origin.  (The Na I D interstellar doublet at $\lambda$5890, not shown in the upper panel of Fig.~\ref{fig:lmc1741}, but quite visible in the lower panel,  is also strong.) LMC174-1 is a member of the Lucke-Hodge 103 OB association (N159/N160), which was studied by \citet{2009AJ....138..510F}. Within the immediate region ($<$1.5') there are four O-type stars: an O9.5~V eclipsing binary OGLE LMC-ECL-22045 with a  {\it B-V} value of $-0.14$, an O6:~V star ([FBM2009] 53) {\it B-V}=+0.26, an O9~V star (FBM2009] 63) with {\it B-V}=$-0.05$, and an O9-9.5~V star (FBM2009] 61]) with {\it B-V}=$-0.03$.  (The photometry all comes from \citealt{Zaritsky04} which has a typical uncertainty of 0.07~mag in $B-V$.)  Comparing these values to that expected for their spectral types (see, e.g., Table 3 in \citealt{Canary}), we find $E(B-V)$ color excesses for these four stars of 0.16, 0.58, 0.26, and 0.28, respectively.  Thus, the reddening in this region {\it is} higher than typical in the LMC, and there is at least one early type star with comparable or higher reddening.   
 
To us, this suggests that the higher reddening is coincidental with the enhanced emission in LMC174-1.  Perhaps the broad emission component to the Balmer lines, stronger He\, {\sc ii} emission, and weaker absorption are instead consistent with a denser stellar wind, and that in some ways LMC174-1 is {\it slightly} intermediate between the WN3/O3 class and normal WN3s. Clearly the star deserves further study.

\section{Completeness}
As detailed above, our survey has uncovered a hitherto unrecognized new class of WR, which we are calling WN3/O3s due to their spectral characteristics.  These stars are generally fainter than the other WRs in the Magellanic Clouds, and with weaker lines.  How do we know that an even fainter and weaker-lined population does not exist?  To answer this, we need to understand exactly what our detection limits have been.

In Figure~\ref{fig:completeness} we show the magnitude difference ($\Delta m$) between the emission-line filter (either {\it WC} or {\it WN}) and the continuum filter ({\it CT}) plotted against the absolute magnitude for all of the LMC/SMC WRs identified in our survey, including those that were previously known.  \citet{MJ98} argue that the detection of WR stars by on-band and off-band interference filter imaging is primarily limited by the emission-line fluxes.  The magnitude differences ({\it WN-CT} or {\it WC-CT}) will be basically proportional to the log of the emission equivalent widths within the on-line band.  However, the signal-to-noise will be higher (and the photometric error smaller) for brighter objects. Consider two intrinsically similar WRs, one of which has a bright companion.  There will be less contrast between the emission-line exposure and the continuum exposure for the binary, but the photometric errors will be less for the binary.  Put another way,  brighter stars with weaker emission lines (as measured by their equivalent widths) may have the same ease of detection as fainter stars with stronger emissions.   Although \citet{MJ98}'s argument applies over most of the parameter space, it breaks down for very weak-lined (in terms of equivalent width) stars with small magnitude differences, as our survey is deep enough so that stars with magnitude differences of only a few hundredths of a magnitude have enough signal-to-noise to be readily detected. Thus for the weakest-lined stars, the equivalent width of the emission matters more than the line fluxes,  as there is a practical lower limit to the photometric error.  This lower limit results when flat-fielding errors and the like begin to dominate over Poisson photon and read noise.  To make this clearer,  we have included in the figure our 3$\sigma$ and  5$\sigma$ photometric limits (shown by the two black curves), where we have set a floor to our photometric errors for {\it WC-CT} and {\it WN-CT} of 0.02~mag.  

There are a number of interesting facts that can be gleamed from Figure~\ref{fig:completeness}.  First, WC stars are indeed much easier to detect than most WNs, a point made by \citet{CM81} and \citet{MC83}, and discussed quantitatively by \citet{MJ98}. (See also \citealt{NeugentM33} and \citealt{NeugentM31}.)   Two tests that are commonly used for evaluating the predictions of stellar evolution theory are the metallicity dependence of the relative number of WC to WN stars (the ``WC/WN ratio") and the relative number WRs to other massive star populations, such as red supergiants (see, e.g.,  \citealt{MasseyARAA,MeynetMaeder05,NeugentM31,BPASS2}).  For such tests to be meaningful, surveys have to be sensitive enough to detect at least most of the WNs.

Secondly, although the WN10-11, slash stars (i.e.,  Ofpe/WN9s), and Of-type stars were challenging to find, our survey had sufficient sensitivity to detect them.  The magnitude differences {\it WN-CT} were only -0.06 to -0.25 for these stars, but thanks to their high optical luminosity ({$V \sim 12$}, or $M_V\sim -7$) they were detected at reasonable significance levels ($>3\sigma$).  All known WN10-11 and Ofpe/WN9 stars were recovered, and we even discovered one previously unknown WN11 star (LMCe063-1), along with numerous Of-type supergiants.   

Third, and most importantly, the ten WN3/O3s\footnote{We are including LMCe055-1 as a member of this class, although the spectrum is more precisely described as WN4/O4.} were not marginal detections.  Even the weakest-lined, faintest of these stars were found at high significance:  LMCe159-1 and LMCe055-1 (the two lowest WN3/O3 points in the plot) were found with significance levels of 10$\sigma$ and 7$\sigma$, respectively, with magnitude differences {\it WN-CT} of -0.22~mag and -0.14~mag.  Were there stars with equally weak lines, we would have detected them at a 3$\sigma$ level even if these were many magnitudes fainter.  Thus, we are confident that there is not a substantial population of weak-lined, faint WRs that we are still waiting to be discovered in the Clouds. 

To put this sensitivity in other terms, a WN3/O3 star like LMC079-1 with a He\,{\sc ii} $\lambda$4686 equivalent width of $-20$~\AA\ and a magnitude difference {\it WN-CT} of  $-0.30$~mag, would still have been detected at a 5~$\sigma$ level if it were two magnitudes fainter ($V\sim 18.3, M_V\sim -0.6$), and at a 3$\sigma$ level they are even more faint ($V\sim 19.1, M_V\sim +0.2$). 

WR stars in binaries will have smaller magnitude differences as explained above, and in extreme cases (when the magnitude difference becomes comparable to the photometric error), the star might not be detected.  For instance, if the WR star were paired with a much brighter companion, such as a RSG (although such a pairing would be unlikely from an evolutionary point of view), the emission would be swamped and we would likely fail to detect the star.  Normal WR binaries, however, with O-type companions were readily found at high significance levels, including two newly found WR+O systems,  LMC173-1 (WN3+O7~V) and LMC143-1 (WN3+O8-9~II)\footnote{Radial velocity studies aimed at determining orbit solutions are underway for these two objects.}.   Even the B0~I+WN composite LH90$\beta$-6 \citep{Waterhouse,Testor93} was found at a high significance level (18$\sigma$).    For us to fail to detect such a binary, a WN star with a typical He~II $\lambda 4686$ equivalent width of $-100$~\AA\ \citep{Conti89} (which would correspond to a magnitude difference {\it WN-CT}$\sim -1.0$~mag) would have to be paired with a star 25$\times$ (or 3.5 mags) brighter. So, for a typical early WN with $M_V$ of $-4.0$ \citep{vanderHucht06}, a companion would have to have $M_V=-7.5$, corresponding to the brightest B-type or later supergiants \citep[see, e.g.,][Table 3] {HumphreysMcElroy}.  For a weak-lined, intrinsically faint WR, such as the WN3/O3 star LMC079-1, a companion would have to be $5\times$ (1.7~mag) brighter for the emission to be lost in the noise.  Thus if it were paired with a star with $M_V=-4.5$, we would probably not have detected it; this would include basically any main-sequence O-type star.    Thus, we cannot rule out the possibility that there are O stars with WN3/O3 companions that we would have missed in our survey\footnote{Would such a pairing be obvious even spectroscopically? The N~V $\lambda \lambda 4603,19$ lines in LMC079-1 have an equivalent width of $-10$~\AA, and if they were diluted by a companion that was 5$\times$ brighter could easily be mistaken for noise. The presence of an WN3/O3 companion with an O star might be only apparent from radial velocity variations, or possibly the UV flux, as the WN3/O3 companion would be significantly hotter (100,000~K vs 30,000-40,000~K).}.

It is reasonable to ask what else we would not have found.  Although we have demonstrated the sensitivity and depth of our survey is sufficient to recover even the weakest-lined, faintest WRs known (plus a new population of previously unknown ones), the seeing at the Swope was often 2\arcsec, which limits our detection in very crowded regions.   For instance, although our image subtraction could show there were at least several  sources in the R136 cluster in 30~Dor, we would certainly not have identified the individual components that were revealed by {\it Hubble Space Telescope} imaging and spectroscopy  \citep{MH98}.

In the Milky Way, numerous central stars of planetary nebulae are hot enough and with sufficient mass-loss rates that their spectra mimic that of classical WR stars; these stars are usually referred to as Pop~II WRs, or [WRs] following  \citet{vanderHucht81}.  These are low-luminosity objects and of course would be undetectable by our survey at the distance of the Magellanic Clouds.  However, halo [WR] members of our own galaxy superimposed on one of the Clouds might be bright enough to be detected.  Many members of this class, though, are of very low excitation WC-type [WC10-11] \citep{vanderHucht81,DePew} and as such would not have the C\,{\sc iii} $\lambda 4650$ emission for which our {\it WC} filter is designed, and the {\it CT} filter would be contaminated by numerous C\,{\sc ii} lines.   This question recently arose due to the discovery of such a $V\sim15.5$ star by Bruce Margon (private comm., 2016) about an object outside of our survey area of the LMC.  Such low-excitation [WC] stars would not be detected, and thus our survey could shed no light on the surface density of such objects.

\section{The Nature of the WN3/O3s: Are They Truly Isolated?}

Here we briefly revisit the evolutionary status of the WN3/O3s.  Nine of the WN3/O3s have remarkably similar spectra, LMCe055-1 being the exception.  \citet{Neugent17} find that they have a nitrogen mass fraction of 0.008 with little or no oxygen or carbon, consistent with the equilibrium products of CNO cycle nuclear burning and an initial metallicity characteristic of the LMC.   Their absolute magnitudes rule out the possibility that they are WN3+O3 binaries, being many magnitudes fainter than an O3~V star by itself. Their bolometric luminosities are similar to other WN3s, but the stars are a bit hotter (see Fig.~2 in \citealt{Neugent17}), leading to their fainter visual magnitudes.   Their mass-loss rates are 3-5$\times$ lower than that of normal WN3 stars (as shown in Fig.~3 of \citealt{Neugent17}), being more like an O-type star.   They have so far only been found in the LMC, and they are not rare there, making up 8\% of the WN population of that galaxy.  No similar stars are known in the Milky Way, suggesting that, whatever their origins, lower metallicity plays a role. 

Additionally, we have reasonable estimates of these stars' {\it masses}. Our modeling give us precise estimates of the effective temperatures.  Combined with the observed SEDs, these temperatures then fix the stellar radii $R$.  The absorption spectrum then allows us to measure the surface gravities, $g$.  Since $g \sim M/R^2$, we can solve for the mass, $M$.  These values range from 6 to 19$M_\odot$, with a median value of 10$M_\odot$.  These, of course, refer to the current masses, and not the initial masses of the progenitors, which presumably were much larger. 

\citet{Neugent17} suggest that these stars are either the result of stripping by a low mass companion, or are a normal stage in the evolution of massive stars at LMC-like metallicities, possibly forming a missing link between O stars and normal WNs.  They favor the latter hypothesis for two reasons: first, because of the metallicity argument (if these formed as a result of binary evolution, why are not similar stars found in the Milky Way?) and secondly because their spatial distribution appears to match that of other WNs in the LMC, as shown in Fig.\ 20 of \citet{Neugent17}.

\citet{SmithWN3O3s} strongly disputes the latter, titling their paper ``Extreme isolation of WN3/O3 stars and implications for their evolution as the elusive stripped binaries."  They compare the distribution of WN3/O3 stars to that of known O-type stars, concluding that they are more isolated than that of other WNs, with a distribution more like that of 8-12$M_\odot$ red supergiants.  This is the same technique used by \citet{2015MNRAS.447..598S} to argue that Luminous Blue Variables are isolated, and hence must be the products of binary evolution.

At first look, their evidence seems irrefutable: there is a very dramatic difference in the separation of the WN3/O3s from the nearest O-type stars compared with the separation of the WNs from the nearest O-type star (see their Figure 1).  For instance, the median separation for the normal WNs is about 0.021 degrees vs 0.2 degrees for the WN3/O3s--a factor of 10 difference!

However, one potential problem with their analysis is that the O-type stellar content of the LMC is poorly known.  Only a handful of star-forming regions have been investigated in detail spectroscopically.  Our experience is that spectroscopy in any star-forming region in the LMC typically uncovers numerous previously unrecognized O-type stars; see, e.g., \citet{Waterhouse}.  \citet{2014yCat....1.2023S} lists about 1500 O-type stars in the LMC with MK types in the literature; of these, the majority have been discovered from recent large-field spectroscopic surveys, such as \citet{2013A&A...558A.134D} and \citet{Evans15}. \citet{SmithWN3O3s} are aware of the potential incompleteness, but argue that the relative comparison with different groups of stars should be valid.  The problem we see with this is that our knowledge of O-type stars is not uniform throughout the LMC, but is concentrated in particularly interesting regions.  With only 10 WN3/O3s known, the effects of small number statistics and the non-uniform sampling of O stars may significantly bias the result.

An additional concern we have with the \citet{SmithWN3O3s} test is that the comparison sample is divided up into WNs with and without hydrogen, not by spectral subtypes.  Studies have shown that WNs of the same effective temperature have similar luminosities whether the stars have hydrogen or not (see, e.g., \citealt{PotsLMC}, Figure 7).  To us it makes more sense to use the sample of early WNs (WN3-WN4s) as the control sample as these stars have similar bolometric luminosities, while the late WNs (WN7, say) have much higher luminosities and may have come from much higher mass stars.  This separation is further supported by considering the distribution of WNs within the LMC, as the WN7s are mostly found in the greater 30~Dor region (Lucke-Hodge associations 90, and 99-100) where few WN3-4s are (Table~\ref{tab:Catalog}).

Let us instead make the comparison using the wide-field photometric survey of \citet{Zaritsky04}.  We restrict the sample to stars with reddening-free indicies $Q=(U-B)-0.72(B-V)<-0.8$, with the additional restriction that $B-V<0$.  $Q=-0.8$ corresponds roughly to that of a B0.5 star, and so this should eliminate stars of B-type and later.  We further restrict the sample to stars with $V<15$, which corresponds to $M_V<-4$ for stars with typical reddening.  This should limit the sample to stars with (initial) masses $>25M_\odot$; see Table 1 in \citet{2017RSPTA.37560267M}.   When we do this we find quite a different result, as illustrated in Fig.~\ref{fig:nathan}: there is very little difference in the separations found between the WN3/O3s and their closest blue star neighbor, and that of the common WN3-4 stars and their closest blue star neighbors!  The median separation between the WN3/O3s and their nearest bright blue neighbor is 156\arcsec, while the median separation between the WN3-4 stars and their brightest blue neighbor is 110\arcsec, no longer a factor of 10 larger.  The {\it Numerical Recipes} \citep{NR} {\sc kstwo} implementation of the Kolmogorov-Smirnov (K-S) test confirms that the cumulative distributions are drawn from the same parent population at the 63\% level.

WR stars with massive binary companions may well have different origins than other WRs, and so we were curious if there would be even better agreement if we eliminated the WN3 and WN4 stars with spectroscopic evidence of OB companions. (There are 37 stars in that sample, down from 62 without this restriction.)  The separation of these stars from the nearest bright blue star is also shown in Fig.~\ref{fig:nathan}.  A K-S test confirms that this matches the cumulative distribution of the WN3/O3s at the 88.5\% confidence level.  The median separation between the nearest bright blue star and the WN3-4 stars without obvious companions is 154\arcsec, essentially identical to the 156\arcsec value for the WN3/O3s.  Based on this we would be hard pressed to agree that WN3/O3 stars are extremely isolated, at least compared with other early WNs in the LMC. 

That said, the comparison sample here is not perfect either: there can be errors in the photometry, and the photometry will be incomplete in areas of high crowding.  Instead let us {\it directly} test if the spatial distribution of the WN3/O3s is similar to that of the normal WN3-4s.  We can do this by measuring the separations between each WN3/O3 stars and the nearest WN3-4 star, and comparing that with the separation between the each WN3-4 star with the nearest other WN3-4 star.  In response to a preprint version of \citet{SmithWN3O3s}, {\citet{Neugent17} added a footnote noting that the median separations were very simpler. Here we expand on the test to include a cumulative distribution, again using the ``single" WN3-4s as our comparison.

The results of this test are shown in Fig.~\ref{fig:nathan2}.    The distributions here are again very similar.  The K-S statistic shows that these are drawn from the same population with a confidence of 74\%.  The lower confidence is due to the fact that there are a few single WN3-4 stars are considerably more isolated from other WN3-4s than the WN3/O3s, i.e., in the {\it opposite sense} that the \citet{SmithWN3O3s} comparison suggests: the WN3/O3s are {\it less} isolated than their common counterparts!

In summary, {\it comparison of the separation of the WN3/O3s from neighboring bright blue stars, and the separations of WN3/O3s from neighboring early WNs, show that the WN3/O3s are no more isolated than that of the other early WN stars in the LMC.}

What does this say about the evolutionary status of the WN3/O3s?   \citet{Neugent17} favored the interpretation that the WN3/O3s are a hitherto unrecognized ``missing link" in the evolutionary process leading to normal WRs in the LMC, where the latter have higher mass-loss rates and less hydrogen than the WN3/O3s, but argues that stripping by a low mass companion is also a viable alternative.    Although our study here contradicts the results of \citet{SmithWN3O3s}, it does not support one interpretation over the other; it simply fails to rule out the ``missing link" scenario.  We do not even consider the ``isolation" question to be completely closed.  For instance, what should we make of the fact that only 20\% of the WN3/O3s are members or near-members of OB associations (Table~\ref{tab:WRs}), while 73\% of the ``single" WN3-4s are found near or in OB associations? (See, e.g., the new WR catalog in the Appendix.)  

\citet{SmithWN3O3s} argue that the solution to this question is multi-epoch spectroscopy, a conclusion with which we strongly concur.   Indeed, we have been engaged in such a practice since our original discovery of the WN3/O3s, with preliminary results suggesting no radial velocity variations as mentioned in \citep{PaperI} and at previous conferences \citep{2017IAUS..329..176N}.  Our most recent observations show that the scatter in our observations is $<$5 km$^{-1}$, with no apparent correlations in the velocities of different lines, but our sample is not yet complete\footnote{\citet{SmithWN3O3s} suggest that higher dispersion data than ours is needed to resolve this question, but the absorption lines in the WN3/O3 stars are rotationally broadened to 120 km s$^{-1}$ \citep{Neugent17}, and so are well sampled at our 3-pixel spectral resolution ($R\sim4100$, or 73 km s$^{-1}$). Even the most narrow emission lines (such N\,{\sc v} $\lambda$4950) have FHWMs of 300 km s$^{-1}$. Thus, going to higher resolution (for example with MIKE) would lead to decreased signal-to-noise (S/N), without improving the radial velocity accuracy.}. Our goal is to obtain $\sim$10 spectra of each of the WN3/O3s, and as shown in Table~\ref{tab:WRs} we are well along on this goal.  The results will be published separately once the data collection is complete, which we expect to be at the next Magellanic Cloud observing season.

\section{Conclusions and Next Steps}

Our four year survey has been remarkably successful and spatially complete. In addition to the 15 WRs found, we discovered 16 interesting emission line objects including 12 Of-type stars, and a LMXB. In this year of the survey we found two rare objects -- both an Of?p and an Onfp. Before our survey, there were only 5 known Of?p stars in the Galaxy and 3 in the Magellanic Clouds \citep{Walborn15}. Amusingly, one of these three stars was discovered by the second author during a previous search for WRs in the SMC \citep{Massey01}. After finding two more of these rare objects in our first year of the survey as well as this third new object during our last year, we have doubled the number of known Of?p stars in the Magellanic Clouds \citep{PaperI}. We have additionally increased the number of known Onfp stars. The current number of Galactic Onfp stars is 17 \citep{Walborn10, Sota11, Sota14, MA16}. Before this survey, there were around 33 members known in the MCs and thus this survey brings the number up to 36 \citep{Walborn10, Walborn14}.

Perhaps the most exciting part of our survey has been the discovery of a new type of WR star, the WN3/O3. These stars have the strong emission lines of WN3s as well as the strong absorption lines of O3Vs. At first glance they might appear to be binaries but they are visually too faint to be WN3+O3Vs. As part of this survey we have found nine of these WN3/O3s in the LMC (and one WN4/O4V), making up $\sim7\%$ of the population of LMC WRs. We have additionally modeled their spectra using {\sc cmfgen} \citep{CMFGEN} and have compared the physical properties with those of more typical LMC WRs \citep{Neugent17}. They have temperatures a bit hotter than average LMC WNs by around 10,000K, putting them at 100,000K. Their abundances are normal. However, their mass-loss rates really set them apart. They are much more similar to that of an O-type star than a WN. At this point we can only hypothesize where these WN3/O3s fit into the evolutionary stage of a WR. A much more detailed look at these WN3/O3s can be found in \citet{Neugent17}. Here we show that unlike the conclusion of \citet{SmithWN3O3s}, they are not isolated objects, at least not when compared to other early-type WNs in the LMC.

As discussed in the section above, we are very interested in any signs of binarity for these WN3/O3s. We are particularly interested in one star, LMCe055-1, which shows an OGLE lightcurve. We have been photometrically monitoring it as well as observing it spectroscopically at quadrature and plan on publishing our results in an upcoming paper. We're additionally monitoring the rest of the WN3/O3s for radial velocity variations. So far we have not seen anything that points to them being in binary systems but we should be able to put more stringent constraints on their binarity within the next year. Hopefully new information about their possible binarity will bring us closer to understanding these strange new WRs.

\acknowledgements
We are dedicating this paper to the memory of our good friend, mentor and colleague Dr. Nolan R. Walborn. As the referee of Paper I, he offered invaluable help in the identification of two of the rare Of?p stars in our sample, and continued to provide advice in classifying peculiar Of stars until the end of this project. In February 2017, when we sent him the spectrum of LMCe136-1 (the new Of?p announced here), his answer was: ``As the principal wizard of the exclusive and secret Of?p order, I do accept this specimen with all its credentials in order." We dearly miss his friendship and advice. Additionally, we are grateful, as always, to the excellent support received at Las Campanas Observatory, and the generous support of both the Carnegie and Arizona time allocation committees for this project over the past four years. We also thank the referee for their helpful comments. We gratefully acknowledge that this research has made use of both the SIMBAD and VizieR catalogue tools. This work was supported by the National Science Foundation through AST-1612874.

\appendix
\section{An Updated Catalog of LMC Wolf-Rayet Stars}

It has been nearly twenty years since the ``Fourth Catalog of Population I Wolf-Rayet Stars in the Large Magellanic Cloud" was compiled (\citealt{BAT99}, hereafter BAT99).  One of the motivations for our survey was the fact that additional WR stars were being discovered serendipitously every few years.  Prior to our survey, \citet{NeugentWO} lists eight additional WR stars that had been discovered since BAT99, along with one demotion.  Since that time our study has discovered another fifteen WRs as detailed earlier in this paper.  In addition, the classification criteria for the ``slash stars" (such as O3If*/WN6) have been refined \citep{2011MNRAS.416.1311C},  and numerous spectroscopic studies have resulted in revised spectral classifications being published for subclasses of the LMC's WRs (e.g., \citealt{2001MNRAS.324...18B}, \citealt{2008MNRAS.389..806S}, \citealt{2003MNRAS.338.1025F}), or clarified the nature of specific objects (e.g., \citealt{2013MNRAS.432L..26S}, \citealt{2017A&A...598A..85S}).

We have taken this opportunity to construct an updated catalog, which we are issuing here as the Fifth Catalog (Table~\ref{tab:Catalog}). In compiling this we were greatly aided by \citet{2014yCat....1.2023S}, an on-line catalog of stellar spectral classifications that contains (nearly) complete literature references. In many cases we were able to re-examine the spectral types, either from the literature  or from our own spectra.  (For the WNs, the on-line resource \citealt{2014yCat..35650027H} [based upon \citealt{PotsLMC}] proved invaluable.)  

We have not updated the catalog of SMC Wolf-Rayet stars, as no new ones have been found since the list given in Table 1 of \citet{2003PASP..115.1265M}.

{\it Non-WRs in the Catalog.}  Although a few of the BAT99 stars are no longer considered WRs, and have been removed from the catalog, others we have included here.    In some cases, the inclusion is due to ambiguity about their revised classification (see, e.g., the note below about BAT99 80).   In other cases, evolution happened.  For instance, as told by \citet{2008ApJ...683L..33W, 2017AJ....154...15W},  the star R127  was classified as an Ofpe/WN9 in the 1970s, but entered a Luminous Blue Variable (LBV) phase in 1982, establishing the Ofpe/WN9 (now often called WN9-11; see \citealt{1997A&A...320..500C}) as a quiescent LBV state. The Ofpe/WN9 star HDE 269582 has now followed suite \citep{2017AJ....154...15W}.  It would seem unsporting to remove these stars from an WR catalog simply because they were temporarily in an outburst state, and indeed BAT99 retained R127 in their catalog (BAT99 83).  It will be most interesting to see if these stars eventually return to an Ofpe/WN9 state.  

{\it Names.}  Although we have provided a running number for our catalog, it is our opinion that the community does not benefit from a new set of designations.  Many WR researchers are already quite used to either the BAT99 numbers or original \citet{1981A&AS...43..203B} numbers, and adding yet another bit of nomenclature seems self-serving. We have included what we believe are the most commonly used names as identifiers; SIMBAD may be checked for additional cross-references.

{\it Coordinates.} Thanks to \citet{2014yCat....1.2023S}, the locations of most of the WRs have been remeasured on the International Celestial Reference System (ICRS).  For the R136 WRs described by \citet{MH98}, we have adjusted their coordinates to the ICRS.  These stars are so crowded that their coordinates are listed by an additional digit.  The coordinates of R136a1 and R136a2 (not observed by \citealt{MH98}) were measured from the same {\it HST} image they used (u2hk0302t), and then transferred as best we could to the ICRS.

{\it Photometry.} \citet{Smith68b} introduced a five-color narrow-band photoelectric photometry system to characterize the continuum flux of WRs, with the bandpasses chosen to exclude (as much as possible) emission lines.  The increased ease of obtaining high quality spectral energy distributions from spectroscopic observations has reduced the usefulness of this system; at the same time, photoelectric photometry has fallen into disuse.   Thus we have chosen to include only the broad-band $V$ measurements rather than the narrow-band $v$ values that were still included in BAT99.  This {\it V}-band photometry does not give as a pristine estimate of the continuum values as did the older {\it v} band, but it does give an idea  of the brightness of the star, and is available for all of the stars in our catalog, not just the ones that were known in an earlier era.   Most of our photometry comes from \citet{Zaritsky04}, although for the most crowded stars, we used other photometry.

{\it Spectral Types.}  Examination of the \citet{2014yCat....1.2023S} catalog of spectral types for the LMC's WRs revealed that most authors were consistent in determining spectral subtypes for the WC stars, but that variations  were quite common for the WNs.  Some of these differences are due to the existence of two different classification schemes for WNs: the ``classical" system first proposed by \citet{Smith68a}, with refinements by numerous authors (e.g.,\citealt{vanderHucht81,1995A&A...293..172C}), and a ``three-dimensional" system proposed by \citet{1996MNRAS.281..163S}.   In the three-dimensional system, the authors added letters to denote line breadth and hydrogen content.   Some have adopted the new system (notably \citealt{2003MNRAS.338.1025F} and \citealt{2008MNRAS.389..806S}), while others have not. \citet{1999NewA....4..489C} argued that the three-dimensional system was (at least) one dimension too many, eschewing the broadness criteria, but concluding there was some usefulness to designating whether hydrogen emission is present.   We are less convinced by the later.  An inspection of Table~2 in \citet{PotsLMC} shows that the classification process for hydrogen is somewhat hit-or-miss, for instance, BAT99 18 and BAT99 23 were both classified as WN3(h) [with the ``(h)" denoting hydrogen emission] by \citet{2003MNRAS.338.1025F}, but modeling showed the actual hydrogen content of the two stars differ significantly: BAT99 18 has a moderately high mass fraction of hydrogen (0.2), while BAT99 23 has no hydrogen.  Similarly BAT99 86 is classified WN3(h) by \citet{2013A&A...558A.134D} but also lacks hydrogen according to modeling of the spectrum \citep{PotsLMC}.  Regardless of the intrinsic merits of the ``three dimensional" system, we note that recent work \citep{NeugentM33,NeugentM31} has classified hundreds of WR stars in nearby galaxies using the older, one-dimensional system, which is more suited to the sort of signal-to-noise that can be readily achieved for stars in galaxies beyond the Magellanic Clouds.   

Here we adopt the specific classical classification criteria summarized in Table 2 of the ``VIIth Catalog of Galactic Wolf-Rayet Stars" by \citet{vanderHuchtII},  but add an ``h" if an inspection of the spectrum shows a strong odd/even Pickering effect indicating the definite presence of hydrogen; i.e., if the even-N He\,{\sc ii} lines (which are coincident with the Balmer hydrogen lines) are strong compared to the odd-N lines\footnote{In other words, we are basically adopting a one-and-a-half dimensional approach to the classification.}. 

Most of the WNs in the LMC are of early (high-excitation) type, i.e., WN3 and 4.  It is worth delineating the classification criteria for these here.   The classification is based upon the relative strengths of the N\,{\sc v} $\lambda \lambda 4603,19$ lines, the N\,{\sc iv} $\lambda 4058$ line, and the N\,{\sc iii} $\lambda \lambda 4634,42$ lines.  For  WN3, N\,{\sc iv}$<<$N\,{\sc v}, and N\,{\sc iii} is weak or absent, while at WN4 the N\,{\sc v}$\approx$N\,{\sc iv} with N\,{\sc iii} weak or absent \citep{vanderHuchtII}.  (For WN4.5s, N\,{\sc iv} is $>$ N\,{\sc v}.) The division between WN3 and WN4 therefore comes down to what one considers ``much less than" as opposed to ``about equal."   We inspected our own spectra of Galactic WN3 and 4 stars, and concluded that for WN4 stars (e.g., HD 4004, CX Cep) the N\,{\sc iv} line really is comparable N\,{\sc v} line in peak intensity, while for the WN3s (e.g., HD 104994, HD211564),  the ratio (N\,{\sc iv} to N\,{\sc v}) is more like 0.1.  We have thus used a ratio of 0.2  and lower to separate the WN3s from the WN4s.   This seems to be different than what some recent authors have done; for instance, \citet{2003MNRAS.338.1025F} clearly favored the WN4 subtype for a number of stars which we call WN3s, i.e., stars with N\,{\sc iv}  less than 20\% as strong as N\,{\sc v} were still considered ``about equal."  We note that our classification is more consistent with older determinations: for instance, both we and \citet{1983ApJ...268..228C} classified BAT99-29, 43, and 59 (for example) as WN3 while  \citet{2003MNRAS.338.1025F} called them WN4s.

{\it Spectral Signatures of Binarity.} For the most part, WRs do not have intrinsic absorption lines.  The obvious exceptions are the transition ``slash stars" \citep{MH98,2011MNRAS.416.1311C} and our newly identified WN3/O3 class.  However, for ``normal" WRs, the presence of absorption was historically taken as evidence of binarity, with the absorption coming from an OB-type companion.  In the late 1970s and early 1980s, this paradigm was challenged by the discovery that the emission and absorption lines moved together in phase in the Galactic WN7 star HD~92740 \citep{1979ApJ...228..206C}, and the lack of radial velocity variations in a number of Galactic WRs such as HD~193077 (WR138, WN5+abs) \citep{1980ApJ...236..526M}, HD~9974 (WR3, WN3+abs), \citep{1981ApJ...244..173M}, HD192641 (WR137, WC7+abs) \citep{1981ApJ...246..145M}.   In some cases, we now know that the absorption comes from distant or line-of-sight companions (e.g., HD~192641 and HD 193077; see \citealt{2016MNRAS.461.4115R}), while in other cases detailed modeling has shown that the absorption is likely intrinsic (HD~9974, see \citealt{2004MNRAS.353..153M}). \citet{Neugent17} discuss other examples, such as the Galactic stars WR7, WR10, WR18, and WR128, and the LMC stars BAT99-18 and BAT99-63.  Modeling has shown that in these cases the absorption are likely intrinsic, and \citet{2009A&A...495..257M, 2013A&A...554A..23M} has argued that such stars may be examples of quasi-homologous evolution. 

The problem is that without radial velocity studies and/or detailed modeling, one can not tell at a glance from the spectrum whether the presence of absorption is intrinsic, evidence of a close binary companion, or is due to a distant or line-of-sight companion\footnote{In the three-dimension classification scheme, \citet{1996MNRAS.281..163S} proposed to distinguish these three cases by using ``ha", ``+OB", and ``(+OB)," with ``+abs" reserved for cases when the origin was not clear.}. Classifying stars {\it post facto} violates the one of the basic tenets of spectral classification; see the exposition in \citet{2011hsa6.conf...25W}.   What we have done here is admittedly somewhat inconsistent: we have generally used the description designed by the reference for the spectral type, but noted in the comments if the star has been shown to be a spectroscopic binary (SB) or not.  Unless otherwise stated, this information is from \citet{2003MNRAS.338.1025F}; however, although most of these stars show absorption lines indicative of a companion, only one of the stars they examined had good enough data to demonstrate that the absorption and emission were moving in anti-phase (SB2).  Additional studies building upon this work the purposes of finding masses would be very worthwhile.

{\it Runaways.} Strong emission lines, such as He\,{\sc ii} $\lambda 4686$, do not give an accurate measure of the absolute radial velocity of the star, due to electron scattering in the wings, as first shown by \citet{1972ApJ...178..175A}.  Furthermore, different emission lines give different measures of the radial velocity of the star, depending upon where they are formed in the outflowing atmosphere.  Thus, identifying ``runaway" WRs, whose radial velocities are discrepant from nearby early-type stars, is hard. \citet{1984PASP...96..968C} identified two such runaways, BAT99 63 and BAT99 81.  \citet{2003MNRAS.338.1025F} disputes that there is anything unusual in the radial velocity of the latter, but offers the identification of seven additional LMC WRs as runaways.  However, they neglected to take into the account the appreciable rotation curve of the LMC. This is a problem; for instance, they identify BAT99 1 and BAT99 2 as runaways because their radial velocities are low compared to the average of all of the other WRs, but these two stars are on the extreme western side of the LMC, where the rotation results in lower radial velocities compared to the average systemic velocity.  Similarly BAT99 157 is described as a runaway because its radial velocity is high compared to the average, but it is at the extreme eastern part of the LMC where the rotation results in a larger radial velocity than the systemic. We have noted these stars designations as potential runaways but list them as uncertain.  More work is needed to identify potential runaways, preferably using lines formed deeper in the photosphere than He\,{\sc ii} $\lambda$4686.

{\it Finding Charts.}  As our coordinates are typically accurate to a fraction of an arcsecond, and modern telescopes point well, we have not seen the need in general to provide finding charts.    There are a few stars for which this could be a problem, and we have provided references to published finding charts in the comments for those stars. 

{\it Facilities:} \facility{Magellan: Baade and Clay (MagE spectrograph)}, \facility{Swope (e2v imaging CCD)}

\bibliographystyle{apj}
\bibliography{masterbib}

\begin{figure}
\epsscale{0.45}
\plotone{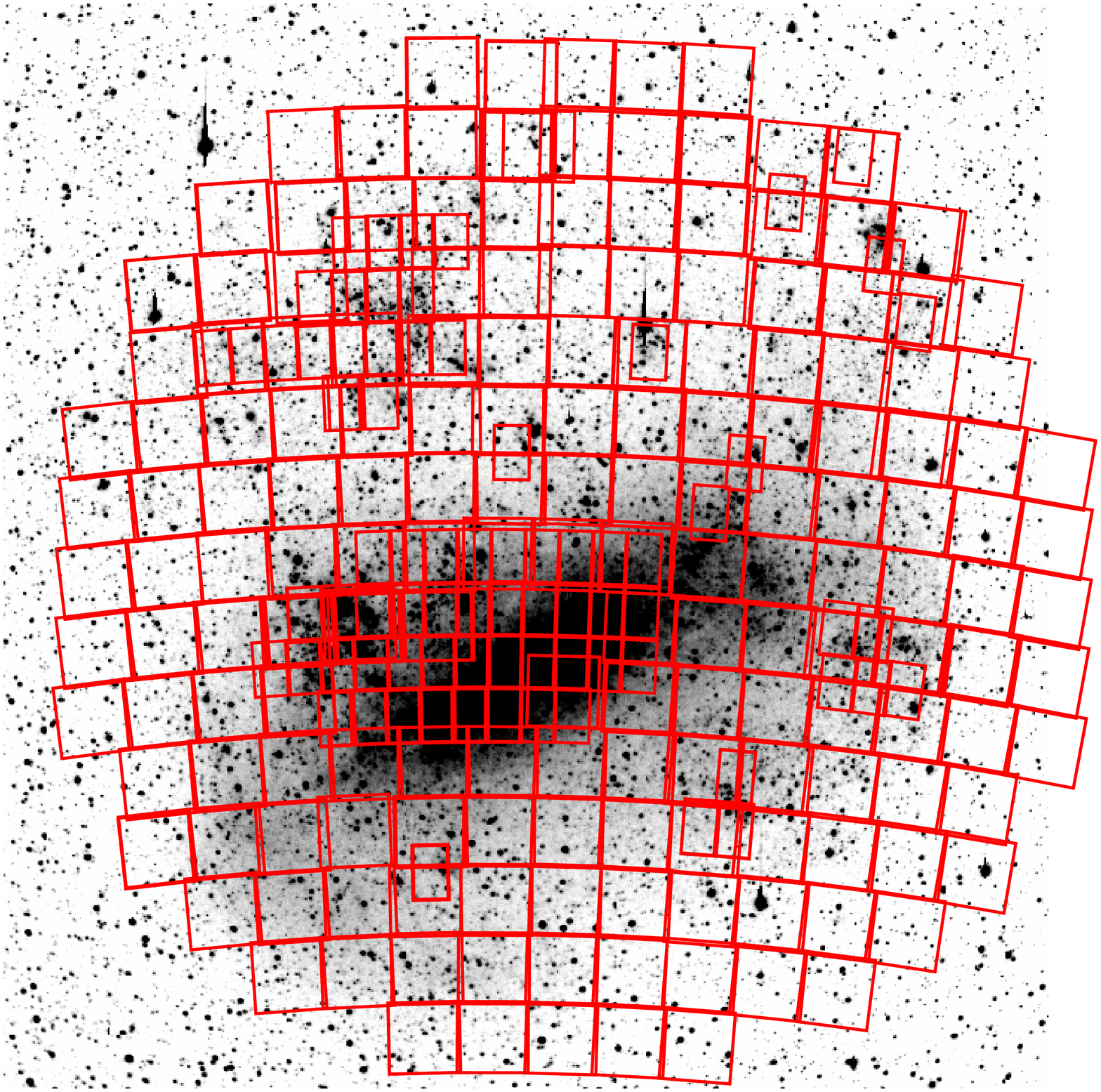}
\plotone{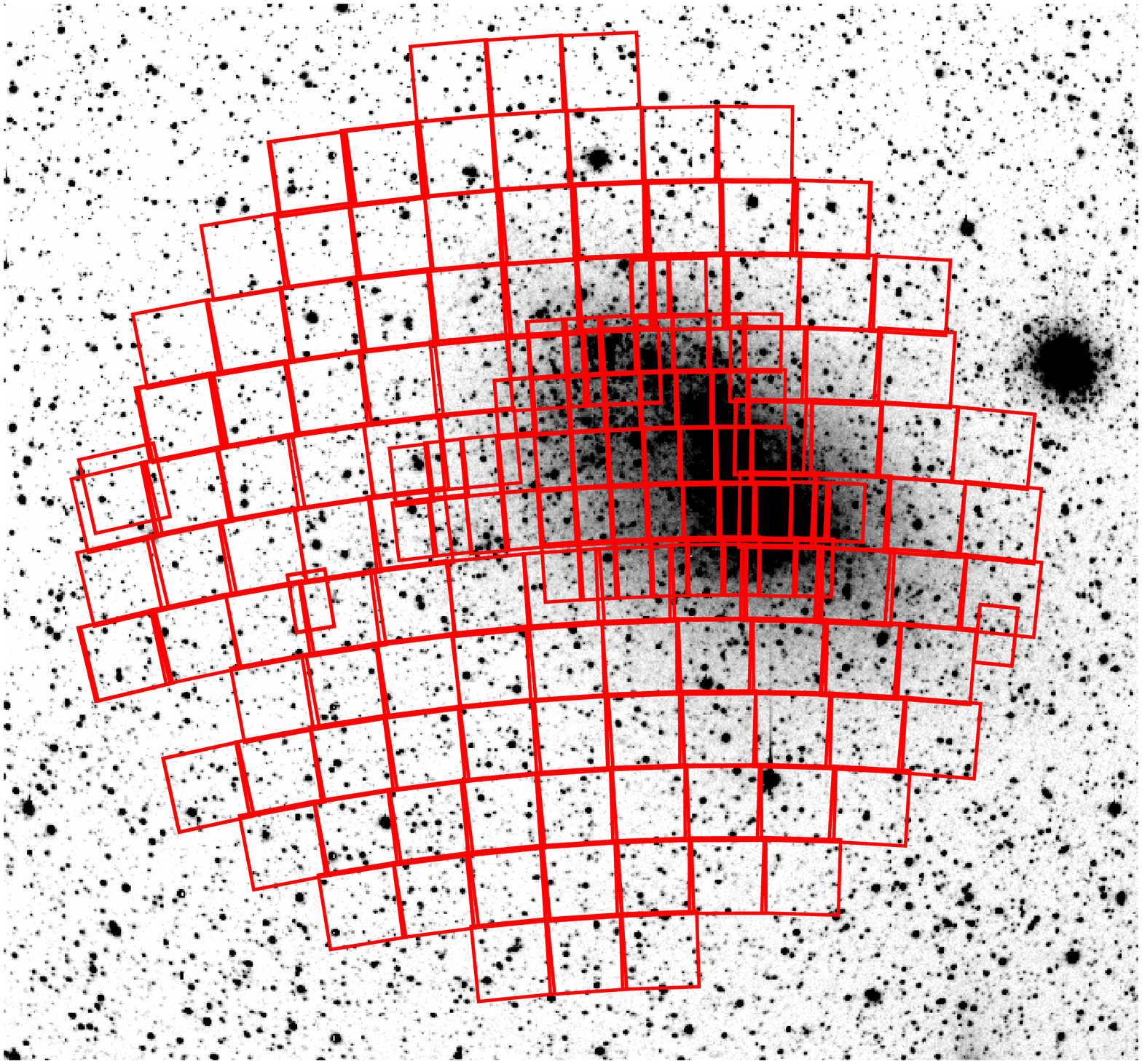}
\caption{\label{fig:coverage} Survey coverage of LMC (left) and SMC (right). Each red box denotes a field observed during our four year survey. There were 178 fields observed in the LMC and 121 fields observed in the SMC. In the LMC, the surveyed region has a diameter of 7$^\circ$, and is centered on $\alpha_{\rm 2000}$ = 5h18m00s $\delta_{\rm 2000}$ = -68$^\circ$45$\arcmin$00$\arcsec$. In the SMC, the surveyed region has a diameter of 6$^\circ$, and is centered on $\alpha_{\rm 2000}$ = 1h08m00s $\delta_{\rm 2000}$ = -73$^\circ$10$\arcmin$00$\arcsec$. Both images come from the R-band ``parking lot" camera of \citet{ParkingLot}.}
\end{figure}

\begin{figure}
\epsscale{0.3}
\plotone{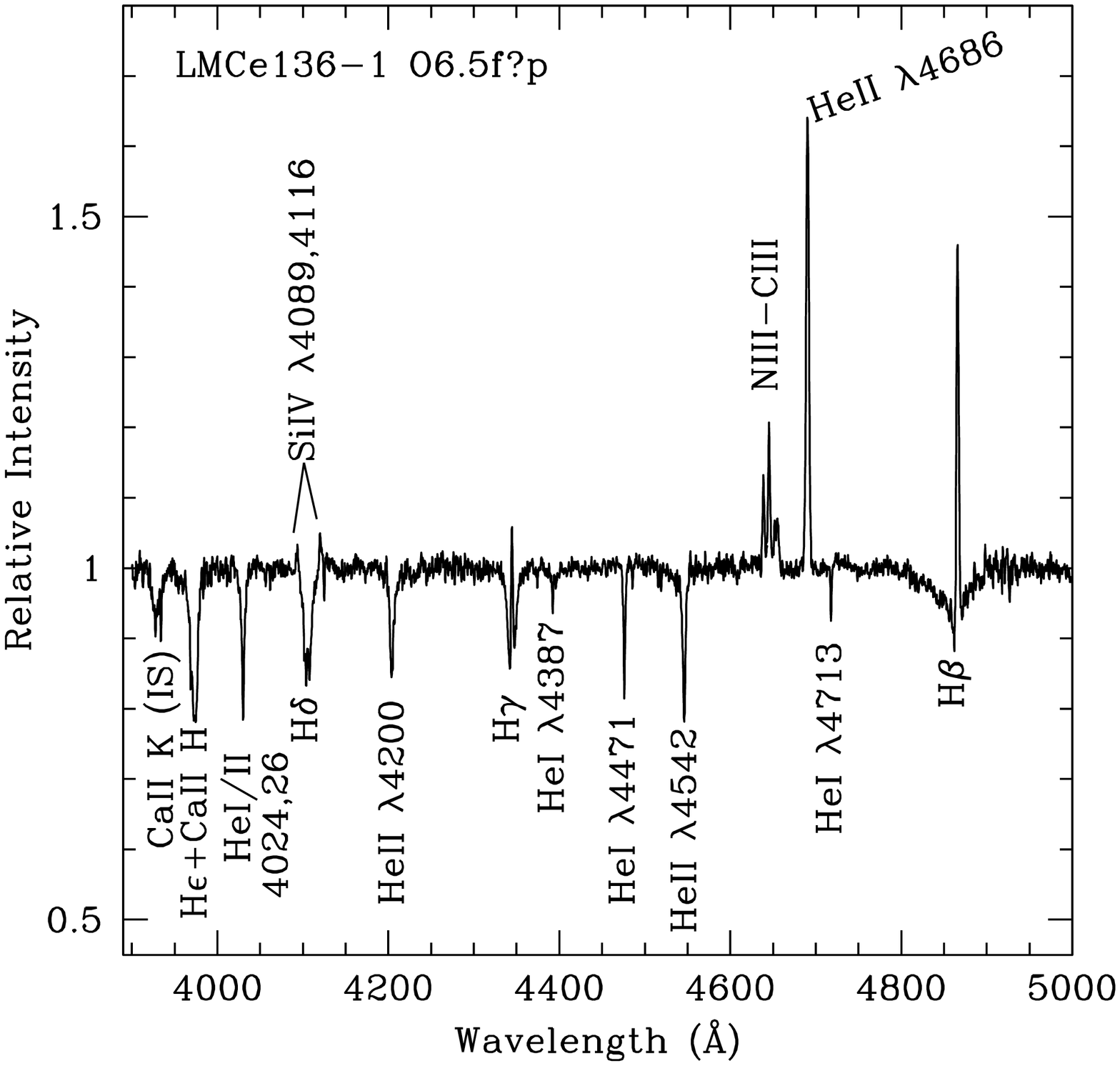}
\plotone{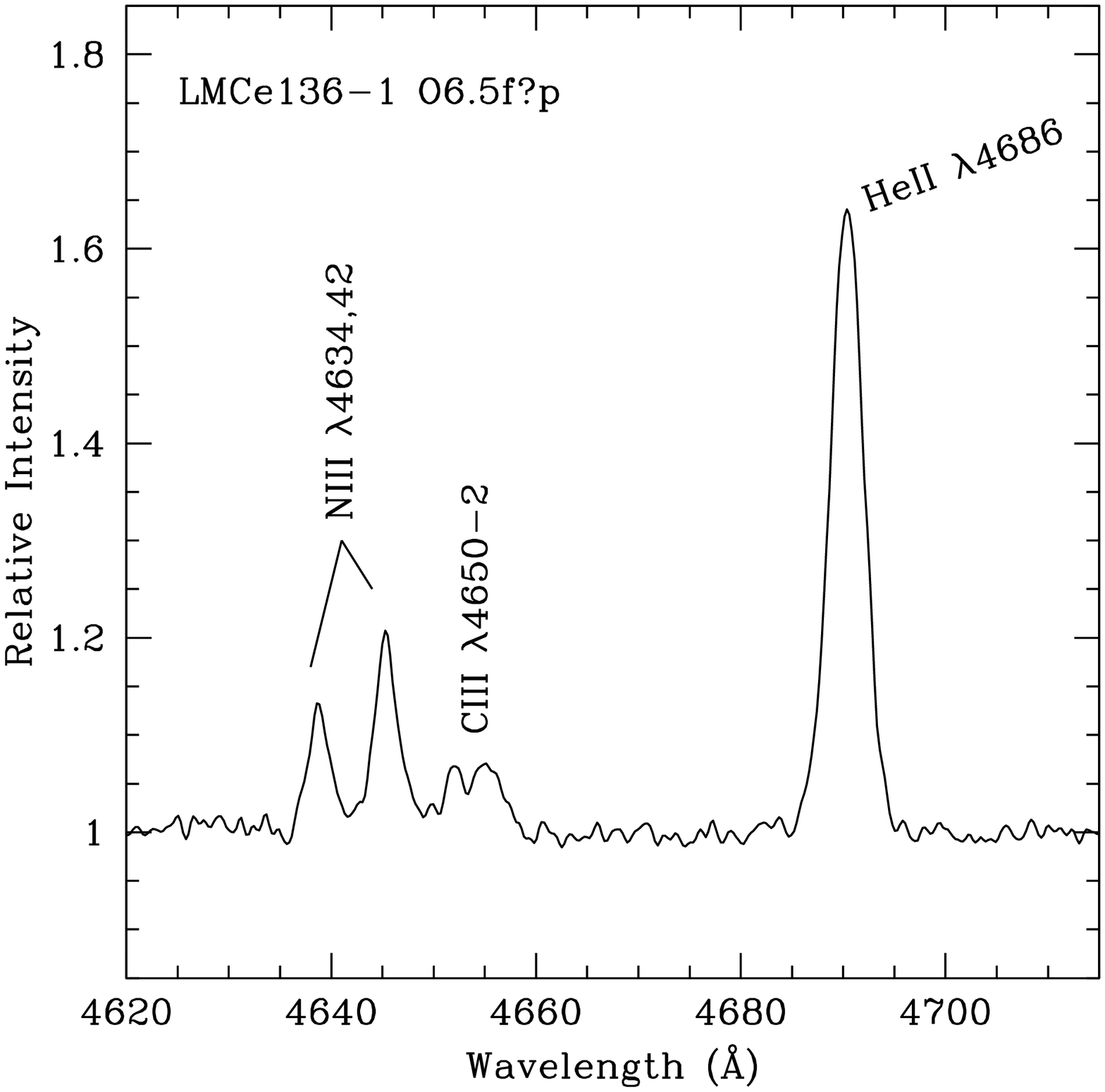}
\plotone{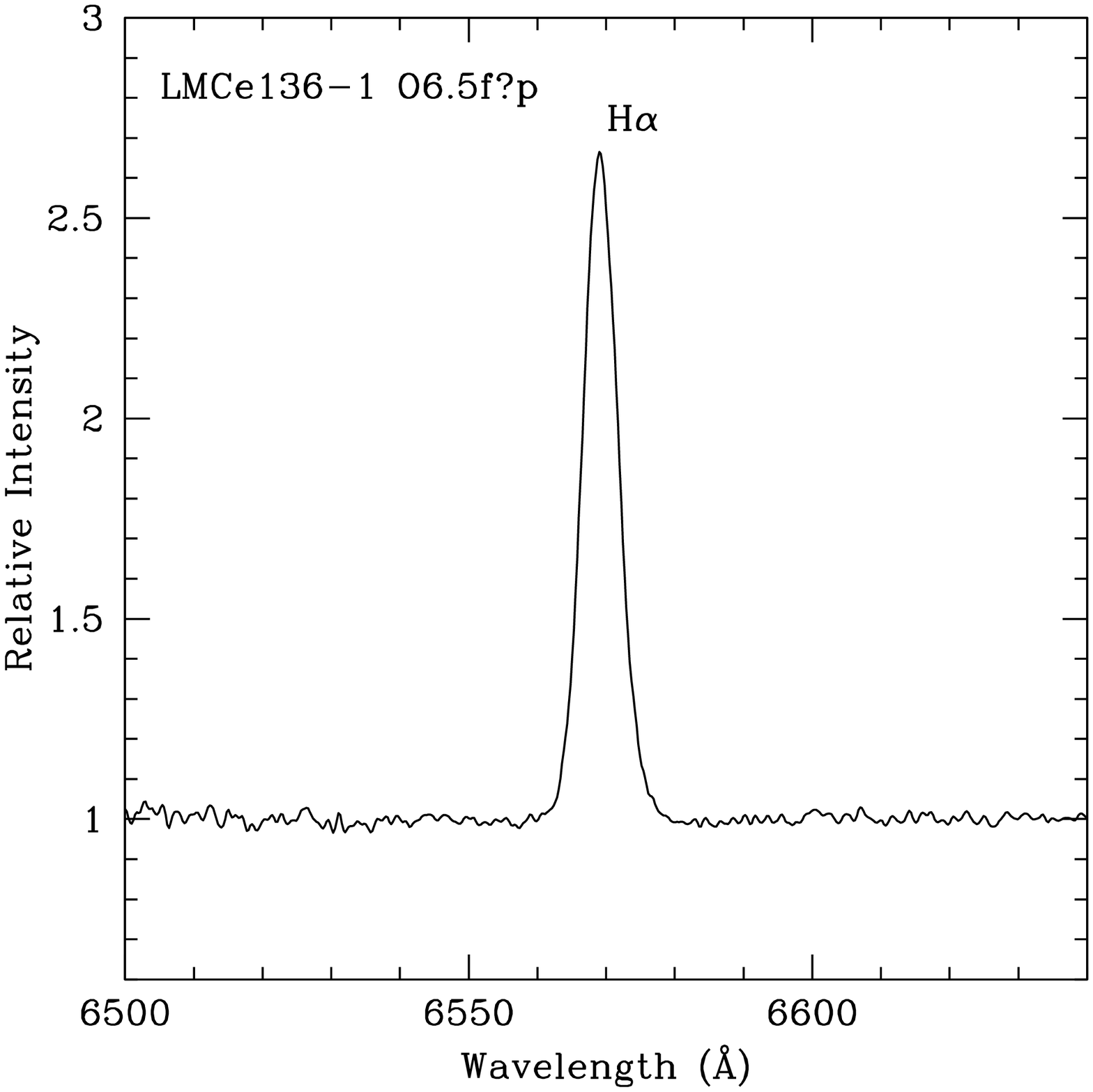}
\caption{\label{fig:Of?p} Normalized spectrum of LMCe136-1, an O6.5f?p star. The O6 classification comes from He\,{\sc ii}$\lambda 4471$ being weaker than He\,{\sc ii}$\lambda 4542$ \citep{WF}. The Of?p classification comes from strong He\,{\sc ii} $\lambda 4686$ emission with much weaker N\,{\sc iii} $\lambda\lambda 4634,42$ and C\,{\sc iii} $\lambda 4650$ emission, plus the presence of emission in the lower Balmer lines.}
\end{figure}

\begin{figure}
\epsscale{0.3}
\plotone{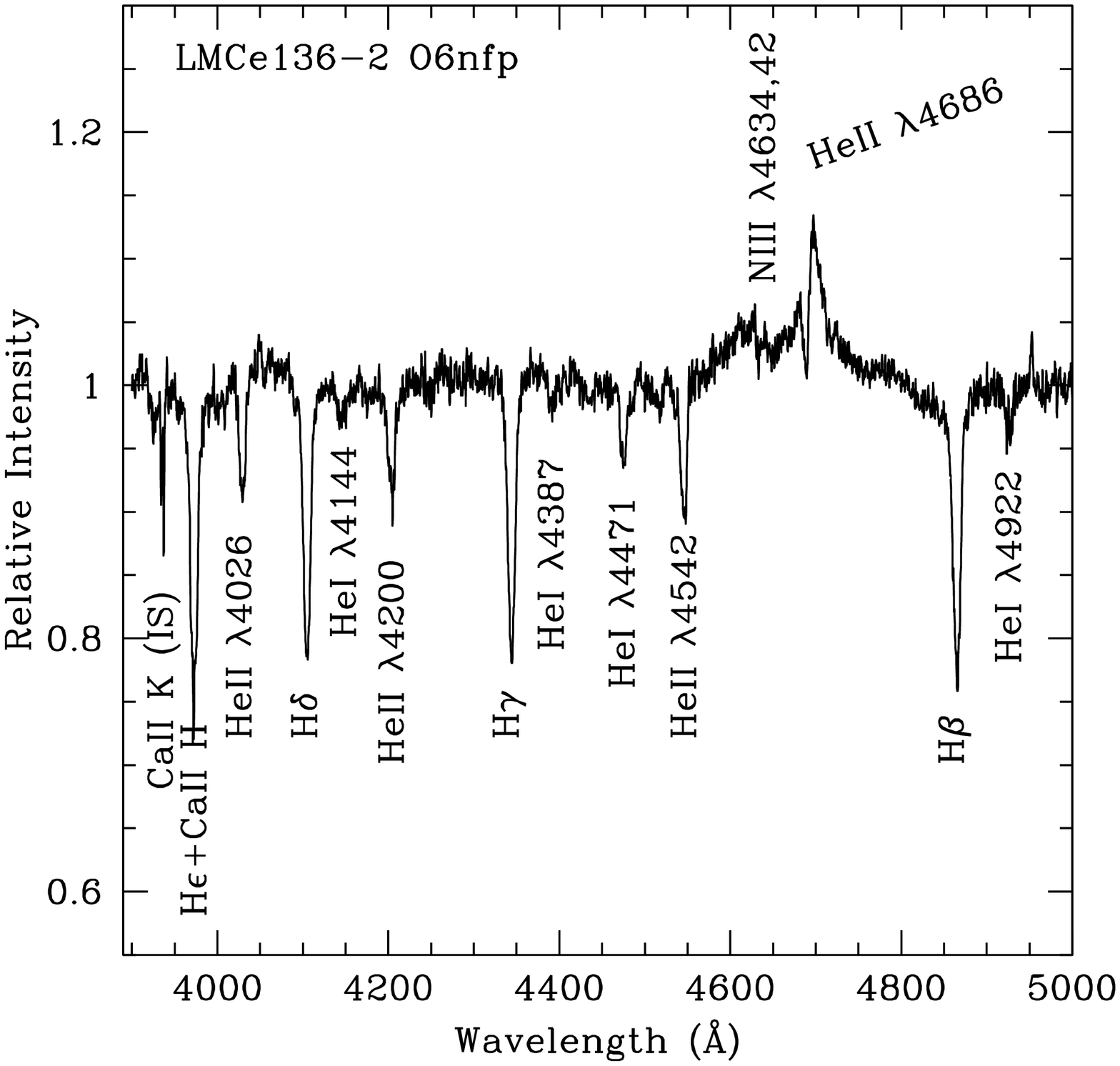}
\plotone{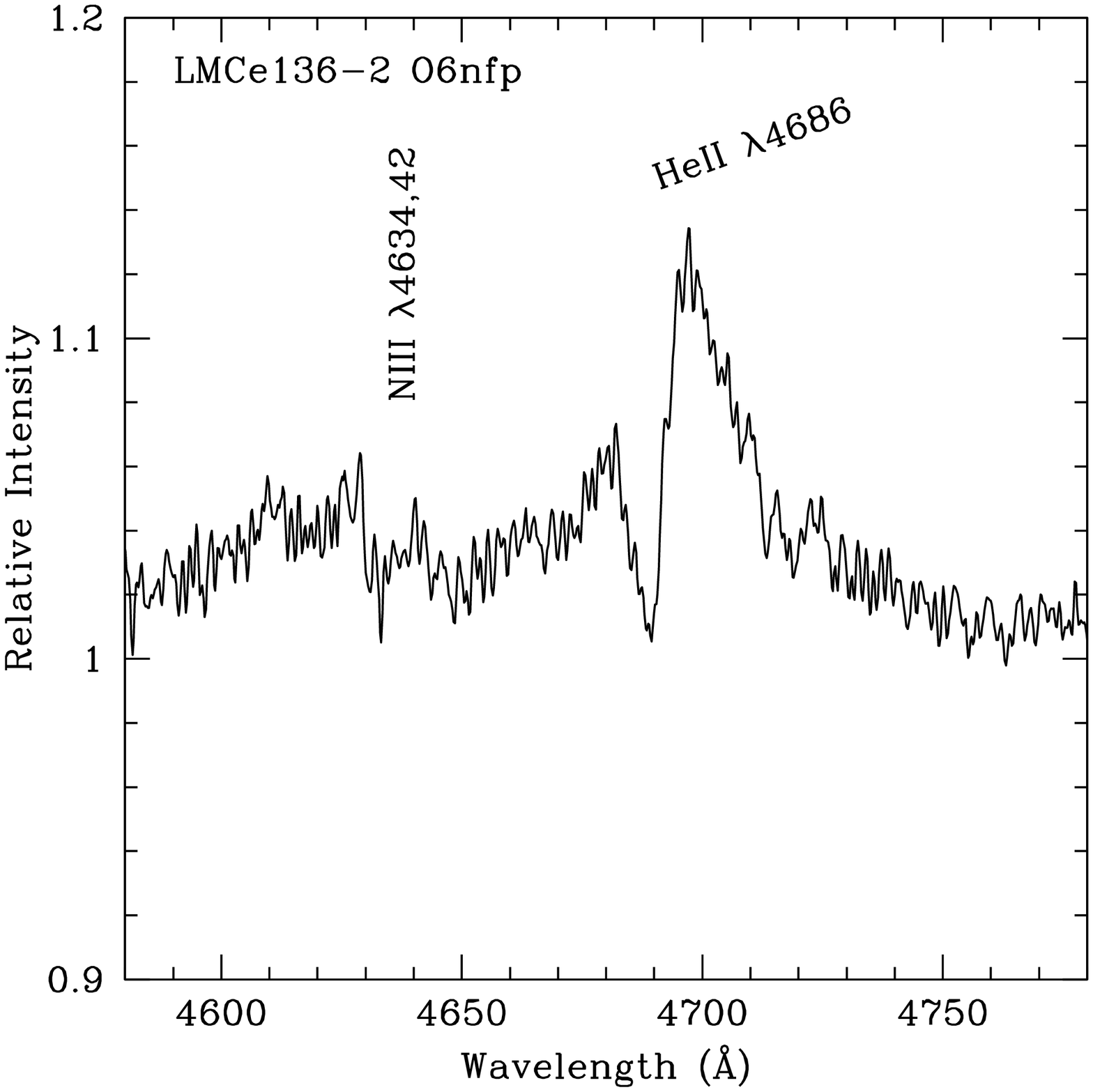}
\plotone{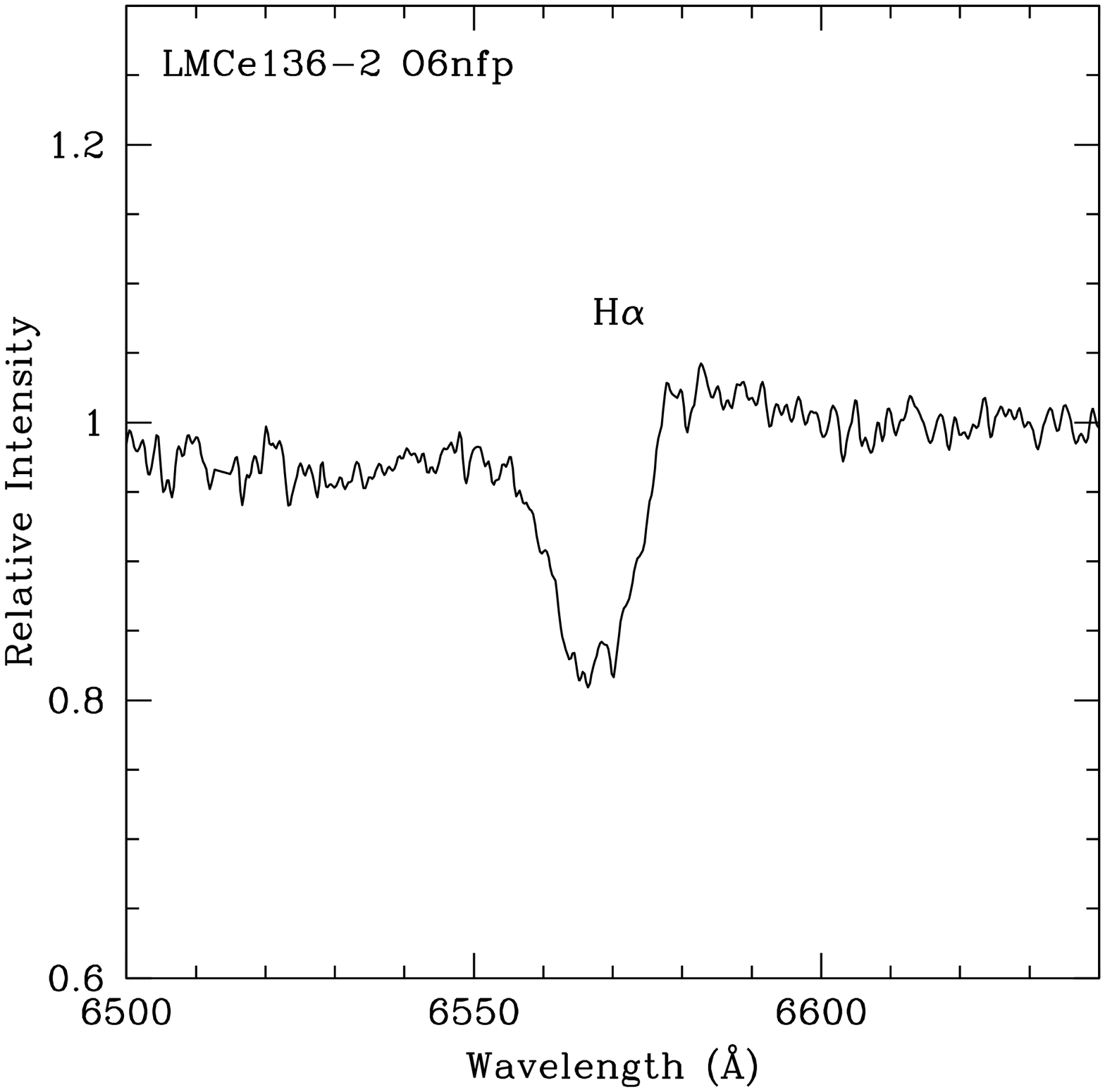}
\caption{\label{fig:Onfp} Normalized spectrum of LMCe136-2, an O6nfp star. The O5 classification comes from He\,{\sc ii}$\lambda 4471$ being weaker than He\,{\sc ii}$\lambda 4542$ \citep{WF}. The Onfp classification comes from He\,{\sc ii}$\lambda 4686$ emission displaying central absorption, broad N\,{\sc iii} $\lambda\lambda 4634,42$ emission, and weak $H\alpha$ \citep{WalbornOnfp}.}
\end{figure} 

\begin{figure}
\epsscale{0.52}
\plotone{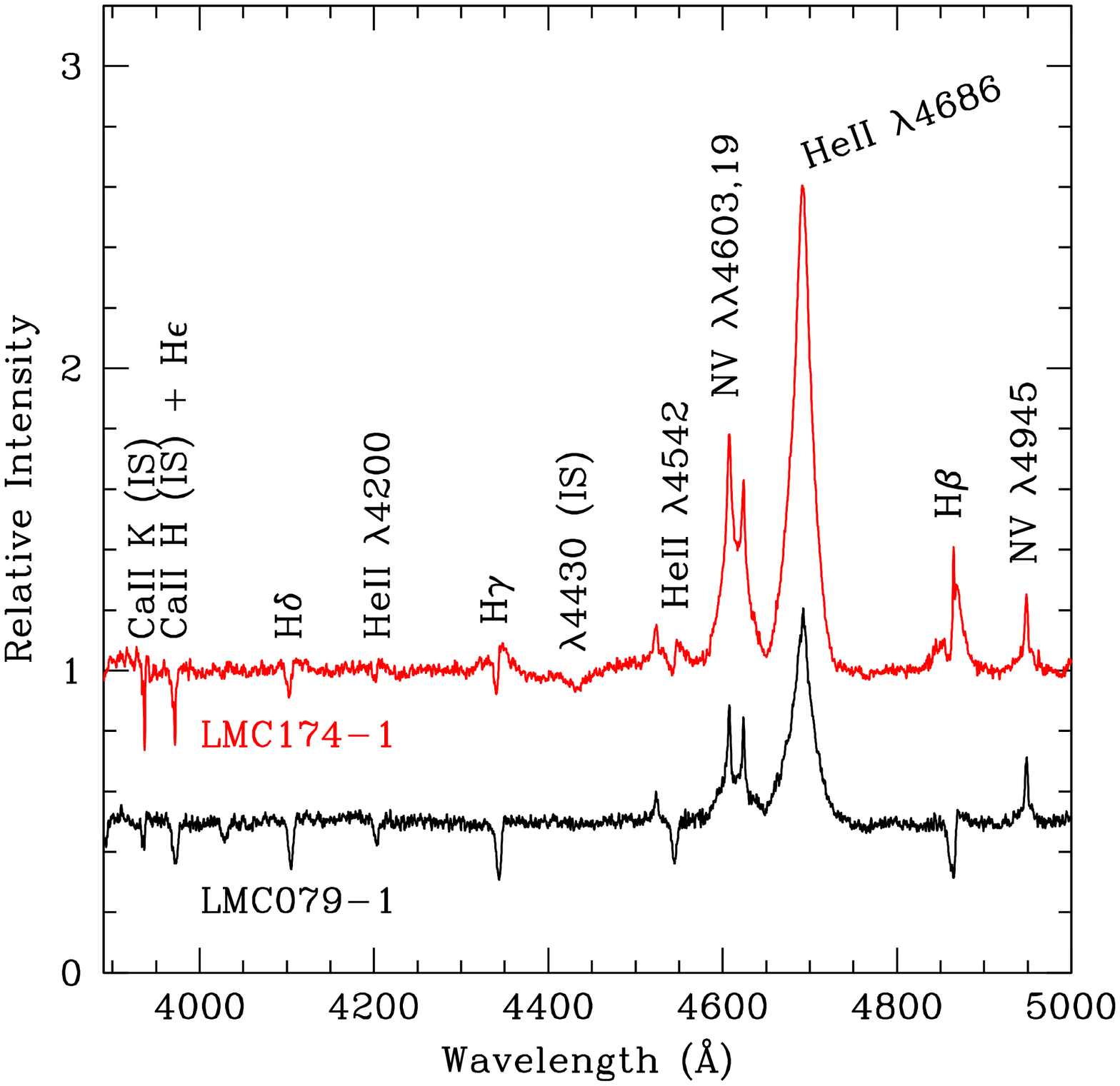}
\plotone{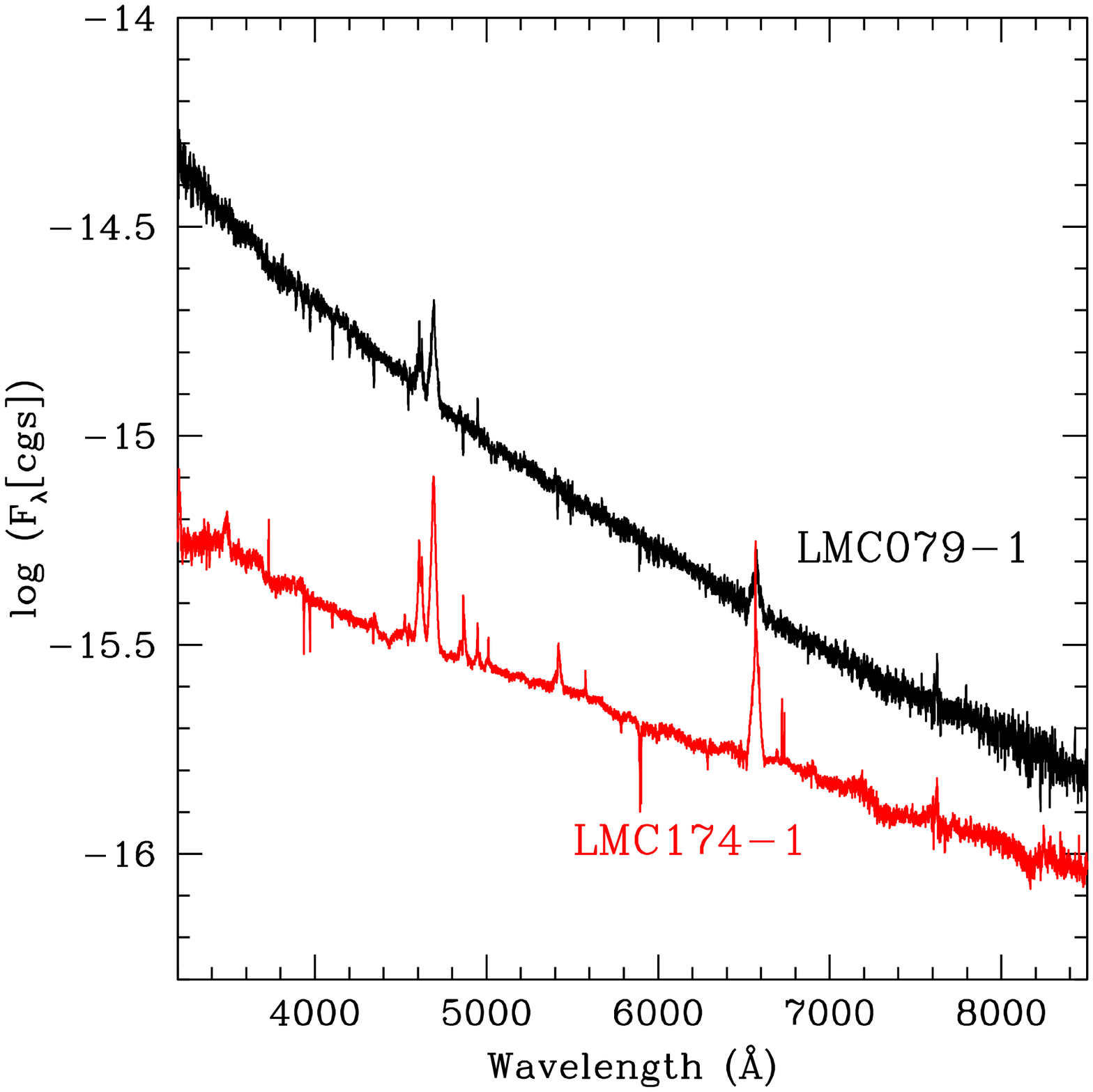}
\caption{\label{fig:lmc1741} LMC174-1.  {\it Upper:} The normalized spectrum of LMC174-1 (red) is compared to that of the more typical WN3/O3 star LMC079-1 (black).  Note that the absorption is weaker, and the emission stronger, in LMC174-1.  The presence of the strong interstellar features, such as interstellar H and K Ca\,{\sc ii} lines and the diffuse interstellar $\lambda$4430 band, suggests higher reddening.  {\it Lower:} The spectral energy distribution of the two stars are compared.  LMC174-1 is significantly more reddened, as indicated by the relatively lower flux at shorter wavelengths.}
\end{figure}

\begin{figure}
\epsscale{1.0}
\plotone{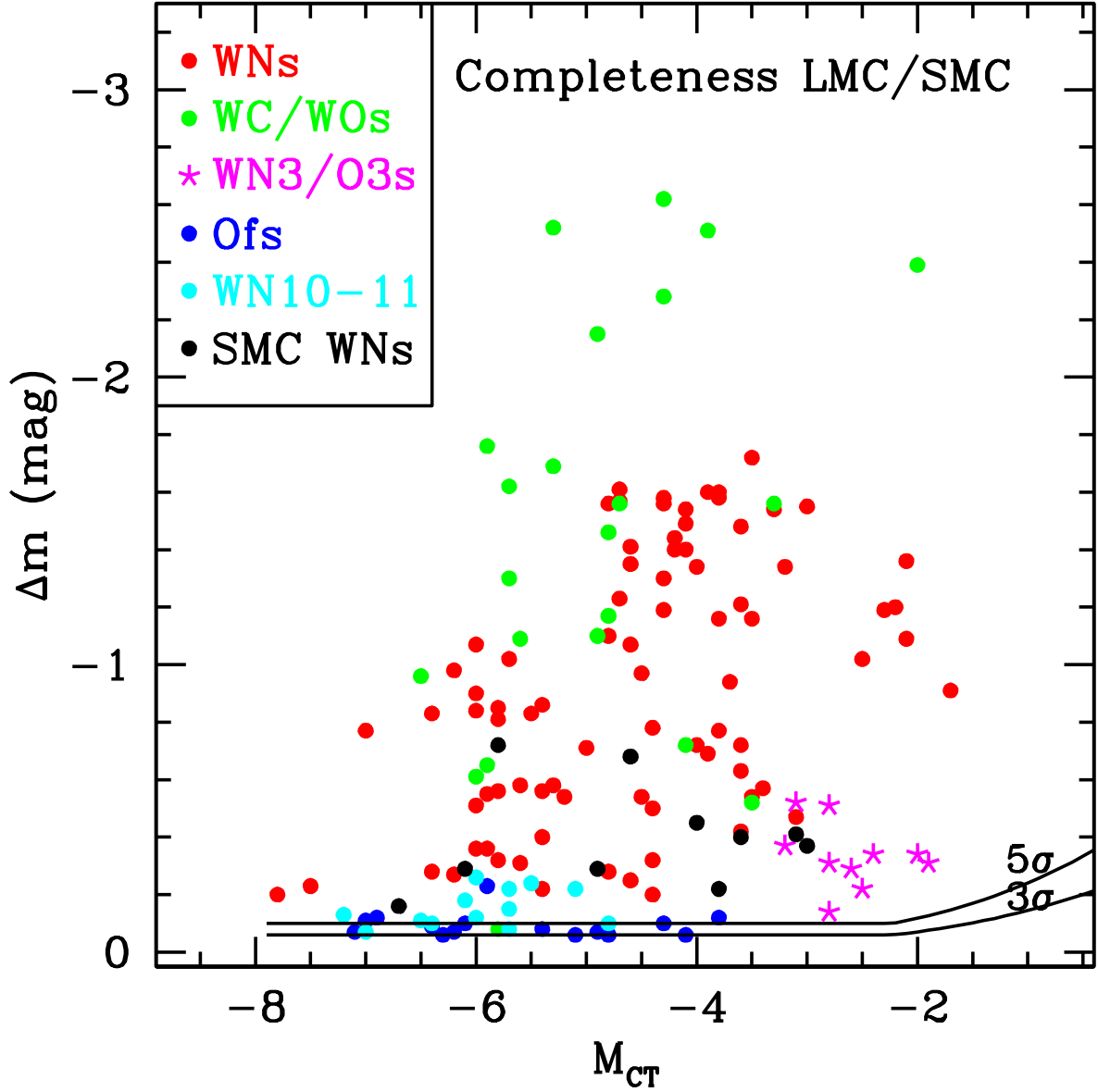}
\caption{\label{fig:completeness} Completeness for our survey.  The magnitude difference {\it WN-CT} or {\it WC-CT} is plotted against the absolute magnitude ($M_{\rm CT} \sim M_{\rm V}$) for all of the SMC and LMC WRs (both known and newly found) in our survey. The slash stars (Ofpe/WN6s) are included with the WN10-11 stars to simplify the plot. We also have included the newly found Of-type stars for comparison.  Our 3$\sigma$ and 5$\sigma$ detection limits are shown as black lines near the bottom of the plot.}
\end{figure}

\begin{figure}
\plotone{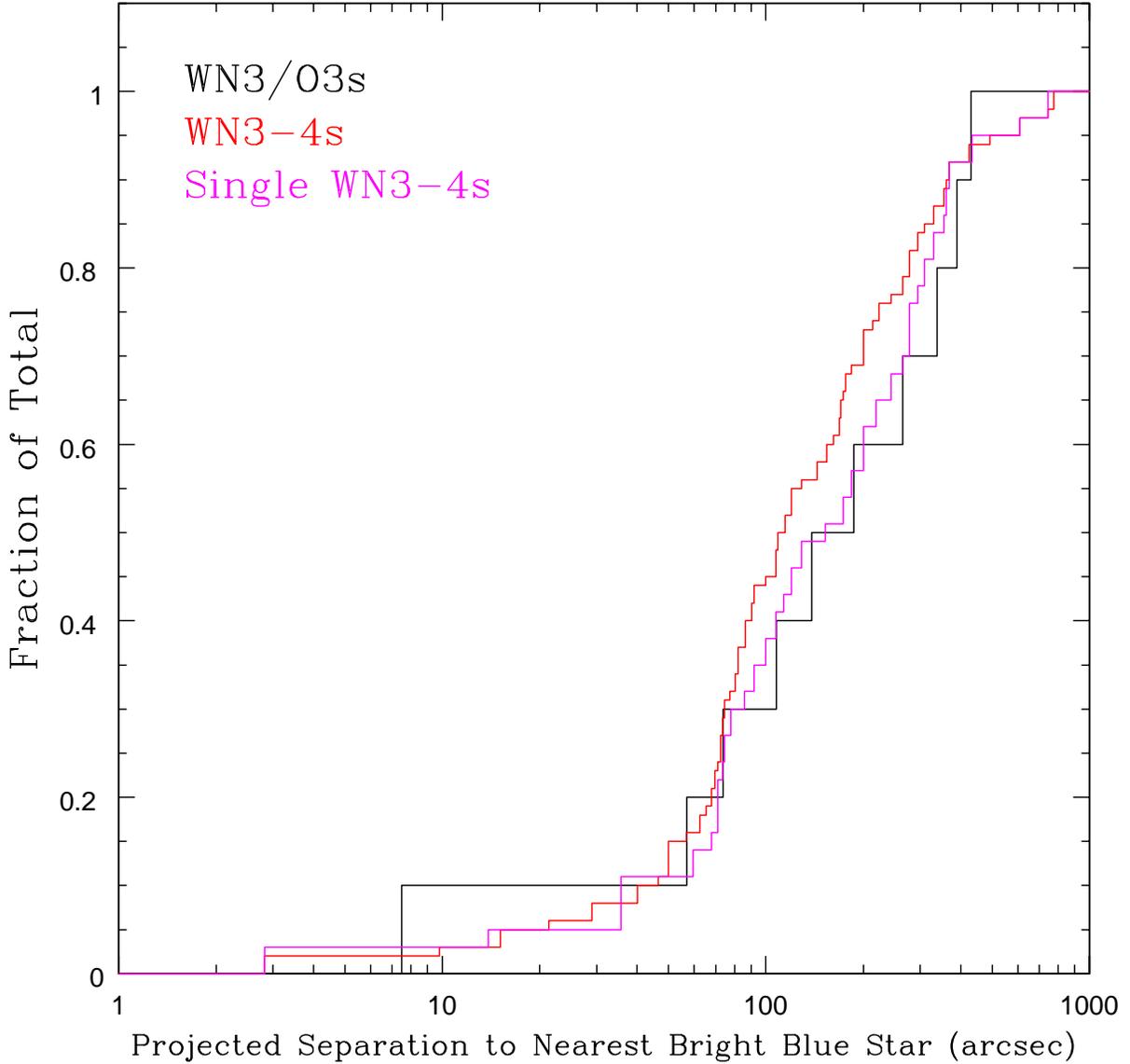}
\caption{\label{fig:nathan} Projected separation from nearest bright blue star.  We show the cumulative fraction of WN3/O3 stars (black) and WN3-4 stars (red) as a function of separation from the nearest bright blue star.  The WN3/O3s show a slightly larger separation from their nearest blue neighbor than do the WN3-4 stars:  for example, 50\% of the WN3/O3s have a separation of 156\arcsec\ or less from their nearest bright, blue neighbor, while 50\% of the WN3-WN4 stars have a separation of 110\arcsec\ or less from their nearest bright, blue neighbor.   If we restrict the latter sample to ``single WN3-4s" (stars without evidence of massive companions in their spectrum, taken as WN3-4s without ``+abs" or ``+OB"), then the distribution with the WN3/O3s is a very good match, as shown by the magenta histogram. The definition of ``bright blue neighbor" is taken to be $V<15.0$, $Q=U-B-0.72(B-V)<-0.8$, $(B-V)<0.0$, which should restrict the sample to stars with initial masses of 25$M_\odot$ or
greater; see text.}
\end{figure}

\begin{figure}
\plotone{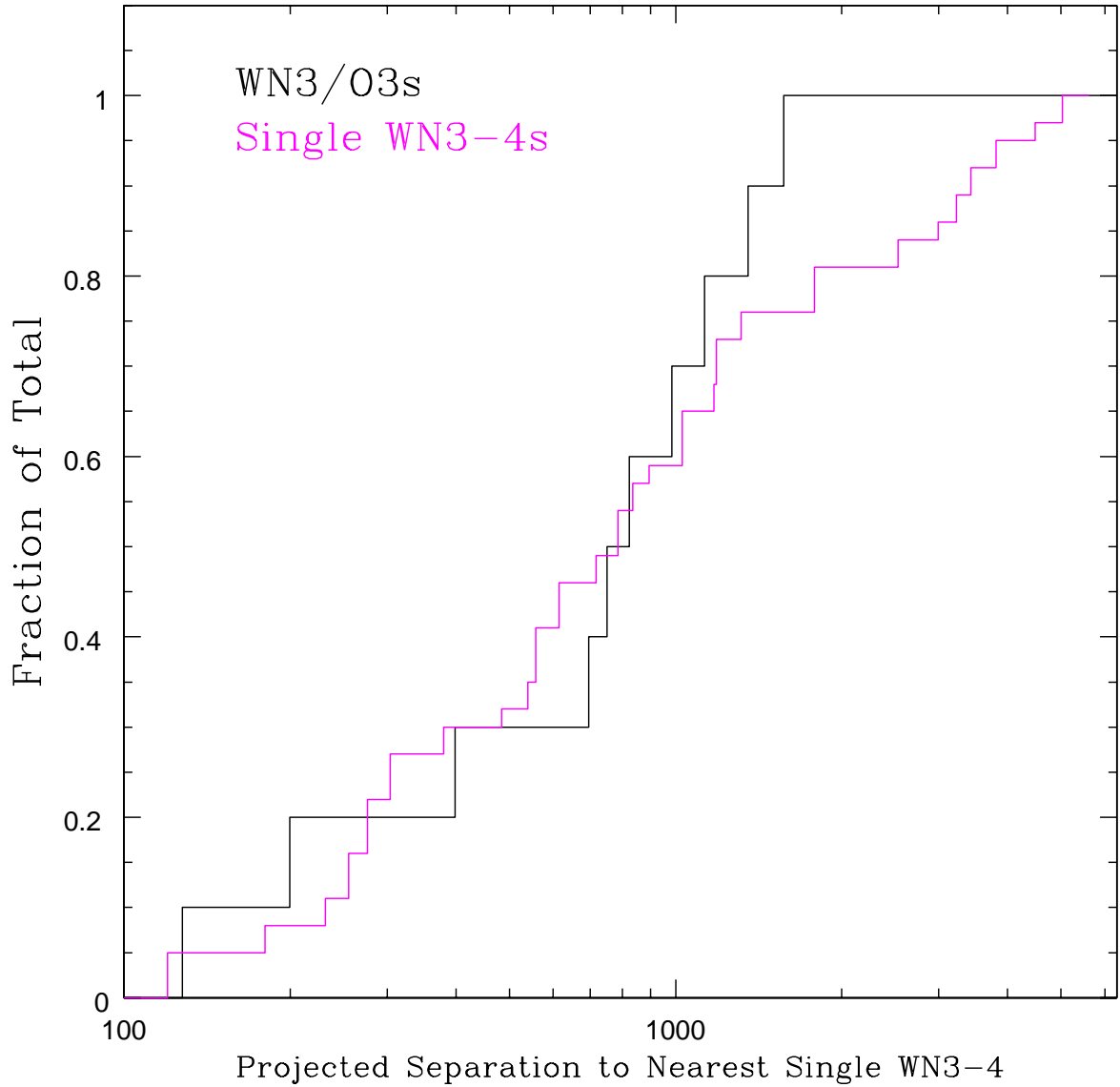}
\caption{\label{fig:nathan2} Projected separation from single WN3-4 stars.   We show the cumulative fraction of WN3/O3 stars (black) and single WN3-4 star (magenta) as a function of the separation to the nearest neighboring single WN3-4 star.  A ``single" WN3-4 star here simply means that there is no obvious sign of a massive companion in its spectrum.}
\end{figure}

\clearpage
\begin{deluxetable}{l l l l l l l}
\tabletypesize{\scriptsize}
\tablecaption{\label{tab:emission} Other Interesting Emission Line Stars}
\tablewidth{0pt}
\tablehead{
\colhead{ID}
& \colhead{$\alpha_{\rm 2000}$} 
& \colhead{$\delta_{\rm 2000}$} 
& \colhead{$V$\tablenotemark{b}}
& \colhead{$B-V$\tablenotemark{b}}
& \colhead{$M_V$}
& \colhead{Sp.\ Type} 
}
\startdata   
LMCe136-1 & 5 32 22.85 & -67 10 18.5 & 14.63 & -0.20 & -4.3 & O6.5f?p\\
LMCe136-2 & 5 33 31.34 & -67 28 57.8 & 14.69 & -0.16 & -4.2 & O6nfp
\enddata
\tablenotetext{b}{Photometry from \citealt{Zaritsky04}.}
\end{deluxetable}

\clearpage
\begin{deluxetable}{l c c c c c c c c c c c c l  l}
\rotate
\tabletypesize{\scriptsize}
\tablecaption{\label{tab:WRs} WRs found as part of this survey}
\tablewidth{0pt}
\tablehead{
\colhead{ID\tablenotemark{a}}
& \colhead{$\alpha_{\rm 2000}$} 
& \colhead{$\delta_{\rm 2000}$} 
& \colhead{$V$\tablenotemark{b}}
&\colhead{$B-V$\tablenotemark{b}}
& \colhead{$M_V$\tablenotemark{c}}
& \colhead{{\it CT}} 
& \multicolumn{2}{c}{\it WN - CT}
&
& \multicolumn{2}{c}{He~II $\lambda 4686$} 
& \colhead{Sp.\ Type} 
& \colhead{Paper}
& \colhead{Comment} \\  \cline{8-9} \cline{11-12}
&&&&& &
&\colhead{mag} 
& \colhead{$\sigma$}
& \colhead{}  
& \colhead{$\log$(-EW)} 
& \colhead{FWHM} 
}
\startdata  
LMC079-1 & 05 07 13.33 & $-70$ 33 33.9 & 16.31   & $-0.25$ & -2.6 & 16.3 &  -0.29 & 13.5 && 1.3 & 30 &  WN3/O3 V     & I & 9 obs \\
LMC143-1 & 05 35 28.52 & $-69$ 40 08.9 & 14.09   & $-0.09$ & -4.8 & 14.1 & -0.28 & 13.8 && 1.2 & 28 &  WN3 + O8-9 III & I & 5 obs \\
LMC170-2 & 05 29 18.19 & $-69$ 19 43.2 & 16.13   & $-0.17$ &-2.8 & 16.1 & -0.31 & 15.6 && 1.3 & 29 &  WN3/O3 V     & I & 9 obs \\
LMC172-1 & 05 35 00.90 & $-69$ 21 20.2 & 15.95   & $-0.12$ & -3.0 &15.9 & -0.46 & 22.1 &&  1.6 & 40 &  WN3/O3 V     & I & 7 obs \\
LMC173-1 & 05 37 47.62 & $-69$ 21 13.6 & 14.65   & $-0.11$ & -4.3 &14.5 & -0.32 & 15.8 &&  1.3 & 29 &  WN3 + O7.5 V   & I & 5 obs \\
LMC174-1 & 05 40 03.57 & $-69$ 37 53.1 & 17.02\tablenotemark{f} & $0.08$\tablenotemark{f} & -3.0\tablenotemark{f} & 17.2 & -0.58 & 17.8 && 1.7 & 30 &  WN3/O3  & I & 2 obs\\
LMC195-1 & 05 18 10.33 & $-69$ 13 02.5 & 15.15 & $0.02$ &-4.1\tablenotemark{g} & 14.8 & -0.72 & 35.9&&  2.6 & 71 &  WO4     &I       & Crowded (LH41, NGC1910)\\
LMC199-1 & 05 28 27.12 & $-69$ 06 36.2 & 16.65   & $-0.22$ & -2.3 &16.5 & -0.34 & 16.0  && 0.2 & 8  &  WN3/O3 V     & I & 4 obs\\
LMC277-2 & 05 04 32.64 & $-68$ 00 59.4 & 15.83   & $-0.16$ & -3.1 & 15.7 & -0.37 & 18.3 && 1.4 & 29 &  WN3/O3 V     &  I & 5 obs\\
LMCe063-1& 05 38 24.21 & $-69$ 29 13.4 & 13.04\tablenotemark{d} &+0.28\tablenotemark{d} & -5.9 & 13.2 & -0.08 & 4.1 && 0.4   & 8 & WN11 & II & Sk$-69^\circ$240 \\
LMCe132-1 &  05 14 17.57 & $-67$ 20 35.1 &  14.34 & $-0.13$ & -5.9 &14.4 & -0.08 & 4.2 &&  0.7  & 8   & O3.5~If*/WN5 & II & \\
LMCe159-1 & 05 24 56.87 & $-66$ 26 44.4 & 16.34 & $-0.23$ & -2.6 &16.4 & -0.22 & 10.2 && 1.3 & 20 & WN3/O3 & II & 3 obs\\
LMCe169-1 & 05 21 22.82 & $-65$ 52 48.8 & 17.12 & $-0.19$ & -1.8 &16.9 & -0.34 & 15.5 && 1.4 & 27 & WN3/O3 & II & 1 obs\\
LMCe078-3 & 05 41 17.50 & $-69$ 06 56.2 & 17.03  & $+0.03$ & -2.2\tablenotemark{e} & 17.2 & -0.27 & 12.0 && 1.2 & 16 & WN3/O3 & III & 1 obs\\
LMCe055-1& 04 56 48.79 & $-69$ 36 40.7 & 16.15 & $-0.10$ & -2.8 & 16.4 & -0.17 & 8.4 && 0.8   & 17 &  WN4/O4 & III & 11 obs; OGLE LMC-ECL-3548 \\
\enddata
\tablenotetext{a}{Designation from the survey.  We have denoted the e2v fields with a small ``e" to distinguish them from our numbering system from \citealt{PaperI}, i.e., LMCe159 is distinct from LMC159.}
\tablenotetext{b}{Photometry from \citealt{Zaritsky04}.}
\tablenotetext{c}{We assume an apparent distance moduli of 18.9 for the LMC, corresponding to a distance of 50 kpc \citealt{vdb00} and an average extinction of $A_V=0.40$ \citep{Massey95, Massey07}.}
\tablenotetext{d}{Based upon our spectrophotometry, the ``emission  free" values would be $V\sim13.11$ and $B-V\sim+0.23$, derived from the continuum fluxes at 4400\AA\ and 5500\AA.}
\tablenotetext{e}{Assumes an extra 0.3~mag of extinction at $V$ given its $B-V$ color.}
\tablenotetext{f}{For all but one of our stars the broad-band colors and fluxed spectra are consistent with the typically low reddening found for OB stars in the LMC $E(B-V)=0.13$, or $A_V=0.40$.  The exception is the visually faintest star in our sample, LMC174-1. \citealt{Zaritsky04} give $V=17.11$ and $B-V=0.14$.   Broad-band photometry of Wolf-Rayet stars is affected by the emission-lines, and so the intrinsic colors are poorly known, but we expect that $(B-V)_0$ for early WNs are similar to that of O stars, around $-0.3$. This would suggest an $E(B-V)\sim 0.44$, or $A_V=1.4.$  Our fluxed spectrum implied $A_V=1.6$, a similarly high value.  As an additional check, we obtained three 300-sec $B$ and three 150-sec $V$ images on the Swope 1.0-m telescope on 2018 Feb 2.  The images were calibrated using isolated stars in common with the \citealt{Zaritsky04} catalog, and we found $V=17.02$ with $B-V=0.08$, only slightly brighter and less reddened ($A_V=1.2$) than the \citealt{Zaritsky04} values.  For computing the absolute magnitude we adopt a compromise value of $A_V=1.4$.}
\tablenotetext{g}{Computed based on the {\it CT} magnitude.}
\end{deluxetable}

\begin{deluxetable}{l l l l l l l c l}                                                                                                                
\rotate                                                                                                                                               
\tabletypesize{\scriptsize}                                                                                                                           
\tablecaption{\label{tab:Catalog}The Fifth Catalog of LMC Wolf-Rayet Stars}                                                                               
\tablewidth{0pt}                                                                                                                          
\tablehead{                                                                                                                                           
\colhead{Num.}                                                                                                                                        
& \colhead{$\alpha_{\rm 2000}$}                                                                                                                       
& \colhead{$\delta_{\rm 2000}$}                                                                                                                       
& \colhead{Preferred Name(s)}                                                                                                                                   
& \colhead{$V$}                                                                                                                                       
& \colhead{Type}                                                                                                                                      
& \colhead{Ref.}                                                                                                                                      
& \colhead{OB assoc.\tablenotemark{a}}                                                                                                                
& \colhead{Comments}                                                                                                                                  
}                                                                                                                                                     
\startdata                                                                                                                                            
  1&04 45 32.25  &-70 15 10.9 &BAT99 1;         Brey 1                         &15.78&WN3             &   1,2,3   &\nodata& Runaway?\tablenotemark{b}                         \\     
  2&04 49 36.25  &-69 20 54.8 &BAT99 2;         Brey 2                         &16.58&WN2             &   2       & (1)  &  N\,{\sc v} present but weak; Runaway?\tablenotemark{b}     \\     
  3&04 52 57.39  &-66 41 13.5 &BAT99 3;         Brey 3                         &14.64&WN3             &   1,2,3   &\nodata&                          \\     
  4&04 53 03.76  &-69 23 51.8 &[M2002]LMC 15666                                &14.39&WN3+O6V         &   4       & (2)  & NErn of 2\arcsec pair    \\     
  5&04 53 30.08  &-69 17 49.4 &BAT99 4;         Brey 3a                        &14.27&WNL/Of?        &   5       & (5)  & E. mem.\ of pair.\ Strong neb.\ WR nature unclear\\     
  6&04 54 28.10  &-69 12 50.9 &BAT99 5;         Brey 4                         &16.74&WN2             &   2,3     &    5  & N\,{\sc v} present but weak.     \\     
  7&04 55 07.60  &-69 12 31.7 &BAT99 5a;        [M2002]LMC  23417              &15.30&WN3             &   7       & (5)  &                          \\     
  8&04 55 31.35  &-67 30 02.7 &BAT99 7;         HD 32109                       &13.85&WN3pec          &   3       &\nodata& Lines unusually broad.   \\     
  9&04 56 02.88  &-69 27 21.5 &BAT99 8;         Brey 8;    HD 32257            &14.23&WC4             &   1,6     &\nodata&                          \\     
 10&04 56 11.0   &-66 17 33.0 &BAT99 9;         Brey 7;    HD 32125            &14.33&WC4             &   1,6     & (10)  &                          \\     
 11&04 56 34.63  &-66 28 26.4 &BAT99 10;      Brey 9;    HD 32228              &10.83&WC4+OB          &   1,6     &    9  &  Crowded\tablenotemark{c}   \\          
 12&04 57 24.10  &-68 23 57.3 &BAT99 11;      Brey 10;  HD 32402               &12.96&WC4             &   1,6     &   12  & SW mem.\ pair             \\          
 13&04 56 48.79  &-69 36 40.7 &[MNM2015]LMCe055-1;  OGLE-LMC-ECL-3548          &16.15&WN4/O4          &   8       &\nodata&                    \\           
 14&04 57 27.44  &-67 39 02.9 &BAT99 12;      Sk$-67^\circ$22                  &13.50&O2If*/WN5       &   9       &\nodata&                    \\           
 15&04 57 41.04  &-66 32 42.6 &BAT99 13;      Sk$-66^\circ$40                  &12.85&WN10            &   10      & (9)  &                     \\          
 16&04 58 56.72  &-68 48 11.0 &BAT99 14;      Brey 11;        HD 268856        &10.83&WN4+OB          &   2,11    &\nodata&                    \\           
 17&04 59 51.55  &-67 56 55.4 &BAT99 15;        Brey 12;      HD 268847        &14.54&WN3             &   12      &\nodata&                      \\  
 18&05 02 59.24  &-69 14 02.3 &BAT99 15a;   [M2002]LMC 57799                   &15.33&WN3+abs         &   7       &\nodata& Candidate WN3/O3       \\       
 19&05 03 08.91  &-66 40 57.5 &BAT99 16;    Brey 13;          HD 33133         &12.70&WN7h            &   2,12    &\nodata& H em.\ present      \\          
 20&05 04 12.33  &-70 03 55.4 &BAT99 17;    Brey 14;          HD 269015        &14.17&WN4             &   1,2,3   &\nodata& WN3?               \\          
 21&05 04 32.64  &-68 00 59.4 &[MNM2014]LMC277-2                               &15.83&WN3/O3          &   13      &\nodata&                     \\          
 22&05 05 08.43  &-70 22 45.1 &BAT99 18;    Brey 15;          Sk$-70^\circ$64  &14.67&WN3h             &   14      & (18)  & H em.\ weakly present \\          
 23&05 07 13.33  &-70 33 33.9 &[MNM2014]LMC079-1;  [M2002]LMC 71747            &16.31&WN3/O3          &   13      &\nodata&                     \\          
 24&05 09 40.42  &-68 53 24.8 &BAT99 19;    Brey 16;          HD 34169         &13.75&WN3+OB          &   1       &   31  &  SB\tablenotemark{b}       \\          
 25&05 09 53.77  &-68 52 52.5 &BAT99 20;    Brey 16a;         [M2002]LMC 81268 &13.74&WC4             &   15      &   31  &                     \\          
 26&05 13 43.77  &-67 22 29.4 &BAT99 21;    Brey 17;          HD 34632         &13.16&WN4+OB          &   2,11    &   37  &                     \\          
 27&05 13 54.27  &-69 31 46.7 &BAT99 22;    Brey 18;          HD 269227        &12.04&WN9             &   10      &   39  &                     \\          
 28&05 13 56.01  &-67 24 36.6 &BAT99 23                                        &17.13&WN3             &   2,16    &   37  &                     \\          
 29&05 14 12.69  &-69 19 26.2 &BAT99 24;    Brey 19;          HD 34783         &14.29&WN3             &   1,2,11  & (35)  &                     \\          
 30&05 14 17.57  &-67 20 35.1 &[MNM2015]LMCe132-1                              &14.34&O3.5If*/WN5     &   17      & (36)  & SWrn of 3\farcs5 pair  \\       
 31&05 14 57.27  &-71 36 18.3 &BAT99 25;    Brey 19a                           &15.11&WN4ha           &   2,14    &\nodata&                    \\           
 32&05 16 38.84  &-69 16 40.9 &BAT99 26;    Brey 20;          Sk$-69^\circ$86  &14.60&WN4             &   1,2,3   & (41)  &                    \\           
 33&05 18 10.33  &-69 13 02.5 &[MNM2014]LMC195-1                               &15.15&WO4             &   13      &   41  & Crowded; FC in Ref.\ 13             \\          
 34&05 18 10.88  &-69 13 11.4 &[L72]LH41-1042                                  &14.00&WO4             &   18      &   41  & Crowded; FC in Ref.\ 13             \\          
 35&05 18 19.21  &-69 11 40.7 &BAT99 27;    Brey 21;          HD 269333        &11.21&BI+WN4          &   2       &   41  & B spectrum dominates  \\        
 36&05 19 16.34  &-69 39 20.0 &BAT99 28;    Brey 22; RMC 89\tablenotemark{d}   &12.21&WC6+O5-6        &   19      &   42  &                     \\          
 37&05 20 44.73  &-65 28 20.5 &BAT99 29;    Brey 23;          Sk$-65^\circ$45  &14.55&WN3+OB          &   1,2,3   &   43  &  SB\tablenotemark{b}                  \\           
 38&05 21 22.82  &-65 52 48.8 &[MNM2015]LMC169-1                               &17.12&WN3/O3          &   17      &\nodata&                     \\          
 39&05 21 57.70  &-65 49 00.3 &BAT99 30;    Brey 24;          Sk$-65^\circ$55  &13.30&WN6h            &   1,2,20  &\nodata& H em.\ strongly present \\      
 40&05 22 04.41  &-67 59 06.8 &BAT99 31;    Brey 25                            &15.21&WN3             &   1,2,3   &   47  &  SB?\tablenotemark{b}                      \\        
 41&05 22 22.53  &-71 35 58.1 &BAT99 32;    Brey 26;          HD 36063         &12.30&WN6h            &   1,2,21  &\nodata& H em.\ present       \\         
 42&05 22 59.78  &-68 01 46.6 &BAT99 33;    HD 269445                          &11.51&Ofpe/WN9        &   22      &   49  &                     \\          
 43&05 23 10.06  &-71 20 50.9 &BAT99 34;    Brey 28;          HD 36156         &12.66&WC4+abs         &   6       &\nodata&                     \\          
 44&05 23 18.01  &-65 56 57.0 &BAT99 35;    Brey 27                            &14.88&WN3             &   1,2,3   &\nodata& H em.\ weakly present  \\       
 45&05 24 24.19  &-68 31 35.6 &BAT99 36;    Brey 29                            &14.37&WN3/WCE+OB      &   1,2,3   &\nodata&                    \\           
 46&05 24 54.34  &-66 14 11.1 &BAT99 37;    Brey 30                            &16.33&WN3             &   2,14    & (52)  &                     \\          
 47&05 26 03.96  &-67 29 57.1 &BAT99 38;    Brey 31;          HD 36402         &11.62&WC4+abs         &   6       &   54  &                    \\           
 48&05 24 56.87  &-66 26 44.4 &[MNM2015]LMC159-1                               &16.34&WN3/O3          &   17      &\nodata& SErn of 3\farcs5 pair  \\        
 49&05 26 30.26  &-68 50 27.5 &BAT99 39;    Brey 32;          HD 36521         &12.32&WC4+O6III/V     &   1,19    &   58  &                     \\          
 50&05 26 36.86  &-68 51 01.3 &BAT99 40;    Brey 33                            &14.77&WN4             &   1,2,20  &   58  &                      \\         
 51&05 26 42.58  &-69 06 57.4 &BAT99 41;    Brey 35;          HD 269549        &14.71&WN4             &   1,2,3   &\nodata&                   \\            
 52&05 26 45.32  &-68 49 52.8 &BAT99 42;    Brey 34;          HD 269546        & 9.86&B3I+WN5         &   14      &   58  & B spectrum dominates     \\         
 53&05 27 37.68  &-70 36 05.4 &BAT99 43;    Brey 37;          Sk$-70^\circ$92  &14.08&WN3+OB          &   1,2,12  &\nodata& SB\tablenotemark{b}    \\          
 54&05 27 42.69  &-69 10 00.4 &BAT99 44;    Brey 36;          Sk$-69^\circ$141 &13.43&WN7h            &   1,2,23  &\nodata& H em.\ strongly  present \\     
 55&05 27 52.66  &-68 59 08.5 &BAT99 45;    HD 269582                          &12.60&WN10$\rightarrow$LBV            &   21,32     &\nodata&                      \\         
 56&05 28 17.90  &-69 02 35.9 &BAT99 46;    Brey 38                            &15.29&WN4             &   1,2,3   &\nodata&                     \\          
 57&05 28 27.12  &-69 06 36.2 &[MNM2014]LMC199-1                               &16.65&WN3/O3          &   13      &\nodata&                      \\         
 58&05 29 12.37  &-68 45 36.1 &BAT99 47;    Brey 39;          HD 269618        &15.30&WN3             &   2,11    &   64  &                      \\         
 59&05 29 18.19  &-69 19 43.2 &[MNM2014]LMC170-2;    [M2002]LMC 143741         &16.13&WN3/O3          &   13      &\nodata&                      \\         
 60&05 29 31.64  &-68 54 28.8 &BAT99 48;    Brey 40;          HD 269624        &14.78&WN3             &   1,2,11  & (64)  &                      \\         
 61&05 29 33.21  &-70 59 34.9 &BAT99 49;    Sk$-71^\circ$34                    &13.48&WN3+O7.5        &   2,24    & (66)  &  SB2\tablenotemark{b,e}  \\         
 62&05 29 53.64  &-69 01 04.8 &BAT99 50;    Brey 41                            &14.52&WN5h            &   2,25    &\nodata& NW mem.\ pair; H em.\ present       \\         
 63&05 30 02.46  &-68 45 18.4 &BAT99 51;    Brey 42                            &15.20&WN3             &   2,3     & (64)  &                    \\           
 64&05 30 12.16  &-67 26 08.3 &BAT99 52;    Brey 43                            &13.56&WC4             &   6       &  \nodata&                    \\           
 65&05 30 38.70  &-71 01 47.8 &BAT99 53;    Brey 44;          HD 37248         &12.96&WC4+abs         &   6       & (69)  &                    \\           
 66&05 31 18.05  &-69 08 45.5 &BAT99 54;    LMC AB 18                          &14.30&WN9             &   21      & (74)  &                     \\          
 67&05 31 25.52  &-69 05 38.6 &BAT99 55;    HD 269687                          &11.87&WN11            &   21      &\nodata&                     \\          
 68&05 31 32.87  &-67 40 46.6 &BAT99 56;    Brey 46;          HD 269692        &14.65&WN3             &   1,2,12  & (76)  & Runaway\tablenotemark{b}  \\          
 69&05 31 34.36  &-67 16 29.3 &BAT99 57;    Brey 45                            &14.90&WN3             &   1,2,3   & (70)  & Runaway\tablenotemark{b}    \\           
 70&05 32 07.49  &-68 26 31.6 &BAT99 58;    Brey 47                            &15.06&WN7h            &   1,2,26  &\nodata& H em.\ strongly present \\      
 71&05 33 10.57  &-67 42 43.1 &BAT99 59;    Brey 48;          HD 269748        &13.19&WN3+OB          &   2,12    & (76)  &   SB?\tablenotemark{b}      \\          
 72&05 33 10.87  &-69 29 01.0 &BAT99 60;    Brey 49                            &14.30&WN3+OB          &   1,2,11  &\nodata&                     \\          
 73&05 34 19.24  &-69 45 10.3 &BAT99 61;    Brey 50;          HD 37680         &13.12&WC4             &   1,6     &   81  &                    \\           
 74&05 34 36.08  &-69 45 36.5 &Sk$-69^\circ$194                                &11.91&B0I+WN          &   27      &   81  & B spectrum dominates  \\        
 75&05 34 37.47  &-66 14 38.0 &BAT99 62;    Brey 51                            &15.31&WN3             &   1,2,11  &\nodata&                      \\         
 76&05 34 52.03  &-67 21 29.0 &BAT99 63;    Brey 52                            &14.60&WN4h            &   1,2,20  & (79)  & H em.\ present. Runaway?\tablenotemark{f}       \\         
 77&05 34 59.38  &-69 44 06.3 &BAT99 64;    Brey 53                            &14.07&WN3+O           &   2       &   81  & SB\tablenotemark{b}  \\  
 78&05 35 00.90  &-69 21 20.2 &[MNM2014]LMC172-1                               &15.95&WN3/O3          &   13      &\nodata& E mem of pair                     \\          
 79&05 35 15.18  &-69 05 43.1 &BAT99 65;    Brey 55                            &15.39&WN3             &   2       & (89)  &  \\   
 80&05 35 28.52  &-69 40 08.9 &[NMN2014]LMC143-1                               &14.09&WN3+O8-9III     &   13      &   87  &                      \\         
 81&05 35 29.80  &-67 06 49.4 &BAT99 66;    Brey 54                            &15.36&WN3(h)          &   2,14    & (84)  &                     \\          
 82&05 35 41.96  &-69 11 52.9 &BAT99 69;    TSWR4                              &16.52&WC4             &   28      &   90  & Crowded\tablenotemark{c} \\                     
 83&05 35 42.19  &-69 12 34.5 &BAT99 67;    Brey 56                            &13.59&WN5h            &   1,2,14  &   90  & H em.\ present  \\              
 84&05 35 42.20  &-69 11 53.6 &BAT99 68;    Brey 58                            &13.59&O3.5If*/WN7     &   9       &   90  & Crowded\tablenotemark{c} \\
 85&05 35 43.49  &-69 10 58.0 &BAT99 70;    Brey 62                            &13.96&WC4             &   6       &   90  &                    \\           
 86&05 35 44.28  &-68 59 36.8 &BAT99 71;    Brey 60                            &14.78&WN3+abs         &   1,2     &   89  &  SB\tablenotemark{b}      \\           
 87&05 35 45.03  &-68 58 44.4 &BAT99 72;    Brey 61                            &15.38&WN4+abs         &   2,3     &   89  & SB?\tablenotemark{b} or candidate WN4/O4      \\        
 88&05 35 50.65  &-68 53 39.2 &BAT99 73;    Brey 63                            &14.64&WN5             &   2,25    &   85  &                      \\         
 89&05 35 52.43  &-68 55 08.7 &BAT99 74;    Brey 63a                           &15.58&WN3+abs         &   2,29    &   89  & Abs strong. WN3/O3 candidate             \\       
 90&05 35 54.03  &-67 02 48.9 &BAT99 75;    Brey 59                            &14.47&WN4             &   2,14    & (84)  &                      \\         
 91&05 35 54.37  &-68 59 07.9 &BAT99 76;    [BE74]381                          &13.27&WN9             &   10      &   89  &                      \\         
 92&05 35 58.87  &-69 11 47.8 &BAT99 77;    Brey 65                            &13.09&WN7             &   30      &   90  & Bright W component; FC in BAT99  \\          
 93&05 35 59.16  &-69 11 50.7 &BAT99 78;    Brey 65b                           &14.59&WN4             &   30      &   90  & Crowded\tablenotemark{c}. HM-5C in Ref.\ 30 \\            
 94&05 35 59.82  &-69 11 22.3 &BAT99 79;    Brey 57                            &13.49&WN7             &   1,2,23  &   90  &                     \\          
 95&05 35 59.89  &-69 11 50.6 &BAT99 80;    TSWR2,N2044W-9A                    &13.02&WN5\tablenotemark{g}        &   2,23,30 &   90  & Crowded\tablenotemark{c}; O4~If$^+$?\tablenotemark{g}  \\                            
 96&05 36 12.13  &-67 34 57.8 &BAT99 81;    Brey 65a                           &15.39&WN5h            &   2,31    &   88  & H em.\ strongly present; Runaway?\tablenotemark{f} WN6h?     \\  
 97&05 36 33.58  &-69 09 17.3 &BAT99 82;    Brey 66                            &16.11&WN3             &   3       &   90  &                     \\          
 98&05 36 43.71  &-69 29 47.5 &BAT99 83;    R127;  HD 269858;                  & 9.30&Ofpe/WN9$\rightarrow$LBV   &   22,32   &   94  &Currently in S Dor state\\         
 99&05 36 51.38  &-69 25 56.7 &BAT99 84;    Brey 68;          HD 38030         &12.99&WC4(+OB)        &   15      &   93  &                      \\         
100&05 36 54.66  &-69 11 38.3 &BAT99 85;    Brey 67                            &12.18&WC4(+OB)        &   15      &   (90)  &                         \\      
101&05 37 11.48  &-69 07 38.2 &BAT99 86;    Brey 69                            &16.30&WN3             &   2,33    & (100)  & \\                              
102&05 37 29.24  &-69 20 47.5 &BAT99 87;    Brey 70                            &13.69&WC4             &   2,6     &   97  & "Close to WO;" see Ref.\ 15\\   
103&05 37 35.72  &-69 08 40.3 &BAT99 88;    Brey 70a                           &16.92&WN3/WCE         &   2       &   99  & CIV 5806 $>$ He\,{\sc ii}; N\,{\sc iv}$<<$ N\,{\sc v} \\  
104&05 37 40.50  &-69 07 57.7 &BAT99 89;    Brey 71                            &14.08&WN6             &   2       &   99  & N\,{\sc v} weak but present  \\         
105&05 37 44.64  &-69 14 25.7 &BAT99 90;    Brey 74;          HD 269888        &14.63&WC4             &   2,15    & (99)  &                     \\          
106&05 37 46.35  &-69 09 09.6 &BAT99 91;    Brey 73                            &14.75&WN6h            &   2,30,33 &   99  & Crowded.\tablenotemark{c} H em.\ strongly present\\       
107&05 37 47.62  &-69 21 13.6 &[NMN2014]LMC173-1;    [M2002]LMC 169271         &14.65&WN3+O7V         &   13      &   97  &                     \\          
108&05 37 49.04  &-69 05 08.3 &BAT99 92;    Brey 72;          HD 269891        &11.62&B1I+WN3         &   23      &  100  & B spectrum dominates   \\       
109&05 37 51.34  &-69 09 46.7 &BAT99 93;    LH99-3                             &13.45&O3If*           &   9       &   99  & Downgraded from O3If/WN6 \\     
110&05 38 24.21  &-69 29 13.4 &[NMN2015]LMCe063-1; Sk$-69^\circ$240            &13.04&WN11            &   17      & (101)  &                      \\         
111&05 38 27.71  &-69 29 58.5 &BAT99 94;    Brey 85;          HD 269908        &14.71&WN3/4pec        &   1,2,3   &  101  & Lines unusually broad\\         
112&05 38 33.62  &-69 04 50.5 &BAT99 95;    Brey 80;          HD 269919        &13.50&WN7             &   2,12    &  100  &                   \\            
113&05 38 36.42  &-69 06 57.4 &BAT99 96;    Brey 81                            &13.74&WN7             &   2       &  100  &  \\
114&05 38 38.84  &-69 06 49.5 &BAT99 97;   [P93]666                            &13.73&O3.5If*/WN7     &   33      &  100  &                      \\         
115&05 38 39.15  &-69 06 21.2 &BAT99 98;    Brey 79                            &13.67&WN6             &   23      &  100  &           \\                    
116&05 38 40.228 &-69 05 59.81&R136-007;    Mk39;            BAT99 99          &13.01&O2.5If*/WN6     &   9,34    &  100  &  Crowded\tablenotemark{h}  \\                          
117&05 38 40.551 &-69 05 57.14&R136-004;    R134; BAT99 100                    &12.89&WN6h            &   34      &  100  & Crowded\tablenotemark{h}   \\                   
118&05 38 41.60  &-69 05 14.0 &BAT99 101;   R140a1                             &12.20&WC4             &   15,35   &  100  & Crowded\tablenotemark{h}                     \\         
119&05 38 41.60  &-69 05 14.0 &BAT99 102;   R140a2                             &12.20&WN6             &   23      &  100  & Crowded\tablenotemark{c}   \\         
120&05 38 41.62  &-69 05 15.1 &BAT99 103;   R140b                              &12.80&WN5(h)+O        &   33      &  100  & Crowded\tablenotemark{c}    \\           
121&05 38 41.875 &-69 06 14.39&R136-044;    BAT99 104                          &14.66&O2If*/WN5       &   9,34    &  100  & Crowded\tablenotemark{h}   \\                          
122&05 38 42.116 &-69 05 55.22&R136-002;    BAT99 105;       Mk42              &12.84&O2If*  &   34,9   &  100  & Crowded.\tablenotemark{h}Downgraded from O3If*/WN6   \\                          
123&05 38 42.333 &-69 06 03.30&R136-006;    R136a3;          BAT99 106         &13.01&WN4.5h          &   34      &  100  & Crowded\tablenotemark{h}                    \\          
124&05 38 42.390 &-69 06 02.95&R136a1;      BAT99 108                          &12.84&WN5h            &   36      &  100  & Crowded\tablenotemark{h}  \\          
125&05 38 42.408 &-69 06 15.04&R136-015;    Mk30                               &13.59&O2If*/WN5       &   9,34   &  100  &   \\                            
126&05 38 42.412 &-69 06 02.89&R136a2;      BAT99 109                          &12.96&WN5h            &   36      &  100  & Crowded\tablenotemark{c}   \\          
127&05 38 42.430 &-69 06 02.75&R136-020;    R136a5;          BAT99 110         &13.93&O2If*/O3If*/WN6 &   9,34    &  100  &  Crowded\tablenotemark{h} \\                           
128&05 38 42.906 &-69 06 04.85&R136-010;    BAT99 112                          &13.47&WN4.5h          &   34      &  100  &  Crowded\tablenotemark{h}  \\          
129&05 38 43.10  &-69 05 46.8 &R136-015;    BAT99 113                          &13.30&O2If*/WN5       &   9       &  100  & Crowded\tablenotemark{h} \\          
130&05 38 43.211 &-69 06 14.40&R136-012;    Mk35;            BAT99 114         &13.54&O2If*/WN5       &   9,34    &  100  & Crowded\tablenotemark{h} \\                           
131&05 38 44.063 &-69 05 55.57&R136-034;    BAT99 115                          &14.47&WC5             &   34      &  100  &  Crowded\tablenotemark{h}            \\                
132&05 38 44.257 &-69 06 05.88&R136-008;    Mk34;            BAT99 116         &13.30&WN4.5h          &   34      &  100  &  Crowded\tablenotemark{h}  \\               
133&05 38 47.52  &-69 00 25.3 &BAT99 117;   Brey 88;          HD 269926        &13.12&WN4.5           &   2       &  100  &  N\,{\sc iv}$>$N\,{\sc v}$>>$N\,{\sc iii}; Runaway\tablenotemark{b}     \\            
134&05 38 53.38  &-69 02 00.9 &BAT99 118;   R144                               &11.11&WN5/6+WN6/7\tablenotemark{i} &   37      &  100  &  WN6h composite type    \\     
135&05 38 55.53  &-69 04 26.7 &[P93]1732;   VFTS682                            &16.08&WN5h            &   38      &\nodata&                     \\          
136&05 38 57.07  &-69 06 05.7 &BAT99 119;   Brey 90;          R145             &11.94&WN6+O3.5If*/WN7 &   39      &  100  &                   \\            
137&05 38 58.09  &-69 29 19.5 &BAT99 120;   Brey 91;          5-68             &12.63&WN9             &   40      &  101  &  \\    
138&05 39 03.78  &-69 03 46.5 &BAT99 121;   [P93]1974                          &15.83&WC4             &   15      &  100  &                     \\          
139&05 39 11.33  &-69 02 01.6 &BAT99 122;   Brey 92;          HD 38344         &13.07&WN5h            &   2,14    &  100  &                     \\          
140&05 39 34.29  &-68 44 09.2 &BAT99 123;   Sand 2;          Brey 93           &15.20&WO3             &   41      &\nodata&                     \\          
141&05 39 36.18  &-69 39 11.2 &BAT99 124;   Brey 93a                           &14.86&WN3             &   2,42    &  103  & NE component; strong neb.            \\       
142&05 39 56.11  &-69 24 24.3 &BAT99 125;   Brey 94;          HD 38448         &13.00&WC4+abs         &   6       &  104  &                     \\          
143&05 40 03.57  &-69 37 53.1 &[MNM2014]LMC174-1                               &17.02&WN3/O3          &   13      &  103  & NW mem.\ of pair                      \\         
144&05 40 07.55  &-69 24 31.9 &BAT99 126;   HD 38472;         Brey 95          &13.17&WN3+O7          &   2,40    &  104  &  SB?\tablenotemark{b,j}     \\           
145&05 40 13.06  &-69 24 04.2 &BAT99 127;   [ST92]4-102                        &13.30&WC4+O           &   15      &  104  &  W mem.\ of pair                   \\          
146&05 40 13.33  &-69 22 46.5 &[MNM2014]LMC174-5;    HD 38489                  &12.22&B[e]+WN?        &   13      &  104  & Broad He\,{\sc ii} $\lambda4686$ em.\\     
147&05 40 50.80  &-69 26 31.8 &BAT99 128;   Brey 96                            &15.02&WN3             &   2,11    & (104)  & Runaway?\tablenotemark{b}       \\       
148&05 41 17.50  &-69 06 56.2 &[MNM2015]LMCe078-3                              &17.03&WN3/O3          &   8       & (111)  &                        \\
149&05 41 48.57  &-70 35 30.8 &BAT99 129;   Brey 97                            &14.70&WN3+O5V         &   43      &\nodata& SB\tablenotemark{b}              \\         
150&05 44 31.03  &-69 20 15.5 &BAT99 130;   Sk$-69^\circ$296                   &12.72&WN11            &   21      &\nodata&                     \\          
151&05 44 53.72  &-67 10 36.2 &BAT99 131;   Brey 98;  Sk$-67^\circ$259         &14.36&WN4             &   2,14    &  116  &                     \\          
152&05 45 24.16  &-67 05 56.8 &BAT99 132;   Brey 99                            &14.62&WN4             &   2,14    & (116)  & Runaway?\tablenotemark{b}                     \\         
153&05 45 51.93  &-67 14 25.9 &BAT99 133;   Sk$-67^\circ$266                   &12.10&WN11            &   21      &  116  &           \\     
154&05 46 46.35  &-67 09 58.3 &BAT99 134;   HD 270149                          &14.50&WN3             &   2       & (116)  & \\
\enddata                                                                                                                                              
\tablecomments{The following stars were included in BAT99 as WRs, but are not included here as they've been reclassified as O-type stars: 
BAT99 6 (now O3f$^*$+O binary,\citealt{2001A&A...369..544N}),  BAT99 107 (now O6.5Iafc+O6Iaf, \citealt{2014A&A...564A..40W}).}
\tablerefs{                                                                                                                                           
(1) Spectrum re-examined here using old SIT-Vidicon spectra described in \citealt{1983ApJ...264..126M} and \citealt{1987ApJS...65..459T};             
(2)  Spectrum re-examined here based upon the spectra shown in \citealt{2014yCat..35650027H};                                                                     
(3)  \citealt{1983ApJ...264..126M};                                                                                                                   
(4)  \citealt{2014MNRAS.442..929G};                                                                                                                   
(5)  \citealt{1991A&A...244L...9M};                                                                                                                   
(6)  \citealt{1986ApJ...300..379T};                                                                                                                   
(7)  \citealt{2012MNRAS.426.1867H};                                                                                                                   
(8)  \citealt{PaperIII};                                                                                                                   
(9)  \citealt{2011MNRAS.416.1311C};                                                                                                                   
(10) \citealt{1995A&A...293..172C};                                                                                                                   
(11) \citealt{1981A&AS...43..203B};                                                                                                                   
(12) \citealt{1983ApJ...268..228C};                                                                                                                   
(13) \citealt{PaperI};                                                                                                                          
(14) \citealt{2003MNRAS.338.1025F};                                                                                                                   
(15) \citealt{2001MNRAS.324...18B};                                                                                                                   
(16) \citealt{1999MNRAS.305..469M};                                                                                                                   
(17) \citealt{PaperII};                                                                                                                          
(18) \citealt{NeugentWO};                                                                                                                   
(19) \citealt{1990ApJ...348..232M};                                                                                                                   
(20) \citealt{1996MNRAS.281..163S};                                                                                                                   
(21) \citealt{1997A&A...320..500C};                                                                                                                   
(22) \citealt{1989PASP..101..520B}                                                                                                                    
(23) \citealt{2008MNRAS.389..806S};                                                                                                                   
(24) \citealt{1995ApJ...438..188M};                                                                                                                   
(25) \citealt{2006A&A...449..711C};                                                                                                                   
(26) \citealt{1992A&AS...93..495M};                                                                                                                   
(27) \citealt{Waterhouse};                                                                                                                            
(28) \citealt{1993A&A...280..426T};                                                                                                                   
(29) \citealt{1985MNRAS.216..459M};                                                                                                                   
(30) \citealt{1999AJ....118.1684W};                                                                                                                   
(31) \citealt{1984PASP...96..968C};                                                                                                                   
(32) \citealt{2017AJ....154...15W};                                                                                                                   
(33) \citealt{2011A&A...530A.108E};                                                                                                                   
(34) \citealt{MH98};                                                                                                                                  
(35) \citealt{1987ApJ...312..612M};                                                                                                                   
(36) \citealt{2010MNRAS.408..731C};                                                                                                                   
(37) \citealt{2013MNRAS.432L..26S};                                                                                                                   
(38) \citealt{2012A&A...542A..49B};                                                                                                                   
(39) \citealt{2017A&A...598A..85S};                                                                                                                   
(40) \citealt{1998A&AS..130..527T};                                                                                                                   
(41) \citealt{1998MNRAS.296..367C};                                                                                                                   
(42) \citealt{1986A&A...162..180H};                                                                                                                   
(43) \citealt{2006A&A...447..667F};                                                                                                                    
}                                                                                                                                                                                                                                                                                                         
\tablenotetext{a}{\phantom{}Lucke-Hodge (LH) OB association numbers are from \citealt{LuckeHodge} and \citealt{LuckePhD}.  Parenthesis are used to denote the association if the star is only near the association.  Note that the 30~Dor region corresponds to LH~100 and its extension to the SW LH~99, while the center of Constellation~III is LH~84.
}                                                                                                                                                     
\tablenotetext{b}{From \citealt{2003MNRAS.338.1025F}.}  
\tablenotetext{c}{Crowded; see finding chart in BAT99.}
\tablenotetext{d}{Note that this star is not HD~35517, but rather its south-preceeding companion, RMC 89.  See \citealt{1987BICDS..33..119B}.}
\tablenotetext{e}{From \citealt{1991IAUS..143..201N}.}
\tablenotetext{f}{From \citealt{1984PASP...96..968C}.}
\tablenotetext{g}{We have retained the WN5 classification of BAT99 80 despite the reclassification by \citealt{1999AJ....118.1684W} as a O4~If+ based upon
an {\it HST} Faint Object Spectrogaph (FOS) spectrum.  Although \citealt{1999AJ....118.1684W}  were able to separate this star ("5C") in the tight cluster that contains Brey 65 and numerous
other early-type stars, there are some problems.  The ground-based spectrum modeled by \citealt{PotsLMC}  shows N\,{\sc iii}, N\,{\sc iv}, and N\,{\sc v} 
emission of roughly comparable strength, as well as He\,{\sc ii} $\lambda$5411 emission.  If this is simply due to contamination by Brey 65, though, where is the N\,{\sc v} emission coming from?  Brey 65 is a WN7 star, and its spectrum has strong N\,{\sc iii}, weak N\,{\sc iv} and no N\,{\sc v}.  So, it is hard to see how the problem is simply contamination. The {\it HST} FOS spectrum is of low S/N and if N\,{\sc iv} and N\,{\sc v}  were present as shown in the ground-based spectrum, we doubt that it would appear.  A new spectrum taken under excellent seeing conditions would be useful in resolving this.}
\tablenotetext{h}{Crowded; see finding charts in BAT99 and \citealt{MH98}.}
\tablenotetext{i}{Described as an SB2 by \citealt{2013MNRAS.432L..26S} but fit with a single set of parameters by \citealt{PotsLMC}.}
\tablenotetext{j}{From \citealt{1998A&AS..130..527T}.}
\end{deluxetable}                                                                                                                                                                                                                                                  

\end{document}